\documentclass[11pt]{article}
\usepackage{graphicx}
\usepackage[cp1251]{inputenc}
\usepackage{rotating}
\begin{document}

 \vskip 0.5cm
  \centerline{\bf\large Milky Way Subsystems from Globular Clusters Kinematics }
  \centerline{\bf\large Using Gaia DR2 and HST Data}
 \bigskip
 \bigskip
  \centerline
 {
  A.~T.~ Bajkova$^1$, G.~ Carraro$^2$, V.~I.~ Korchagin$^3$, N.~O.~Budanova$^3$, V.~V. ~Bobylev$^1$
 }
 \bigskip
 \centerline{\small \it
$^1$Pulkovo Astronomical Observatory, St.-Petersburg, Russia}

 \centerline{\small \it
$^2$Dipartimento di Fisica e Astronomia, Universit\'a di Padova, Vicolo Osservatorio 3 I-35122, Padova, Italy}

\centerline{\small \it
$^3$Southern Federal University, Rostov-on-Don, Russia}

 \bigskip
 \bigskip

 \begin{abstract}
We employ Gaia DR2  proper motions for 151 Milky Way globular clusters from Vasiliev (2019) in tandem with distances and line-of-sight velocities to derive  their kinematical properties. To assign clusters to the Milky Way thick disk, bulge, and halo we follow the approach of Posti et al. (2018)  who distinguished among different Galactic stellar components using stars's orbits. In particular, we use the ratio $L_{z}/e$,  the $Z$ projection of the angular momentum to the eccentricity, as population tracer, which we complement with chemical abundances extracted from the literature and Monte-Carlo simulations.
We find that 20 globular clusters belong to the bar/bulge of the Milky Way, 35 exhibit disk properties, and 96 are members of the halo. Moreover, we find that halo globular clusters have close to zero rotational velocity  with  average value  $<\Theta>$ =1$\pm$ 4 km s$^{-1}$. On the other hand, the sample of clusters that belong to the thick disk possesses a significant rotation with average rotational velocity 179 $\pm$ 6 km s$^{-1}$. The twenty globular clusters orbiting within the bar/bulge region of the Milky Way galaxy have average rotational velocity of 49 $\pm$ 11 km s$^{-1}$.

\end{abstract}


\bigskip

 \section{INTRODUCTION}
The study of kinematical properties of the Milky Way globular clusters (GCs) has a long tradition. One of the most important step in understanding the nature of the globular cluster system was done by Zinn (1993) who suggested a cluster classification based on the horizontal branch morphology and chemical abundances. Zinn (1993) found that the metal rich globular  clusters ($[Fe/H]> -0.8$) are confined to the bar/bulge and to the disk of the Galaxy, while the metal-poor globular clusters ($[Fe/H]\leq -0.8$) are generally located in the Galactic halo. Later on, Mackey $\&$ Gilmore (2004) used Zinn's (1993) classification and compared the cluster properties with those in the Large and the Small Magellanic Clouds concluding that all young globular clusters and 15 -17 percent of the old ones are of external origin.

More recent progresses in understanding the structural properties of the Milky Way galaxy come from the  Gaia space mission. The recently released Gaia DR2 catalog Gaia Collaboration (Helmi et al., 2018) allowed to make significant progress in understanding the nature and the origin of the Milky Way globular cluster system. Gaia collaboration (Helmi et al., 2018) determined proper motions of seventy five globular clusters. In the subsequent paper Posti $\&$ Helmi (2019) used the Gaia Data Release 2 proper motions catalog of
75 globular clusters and  measurements of the proper motions of another 20 distant clusters obtained with the Hubble Space Telescope, and determined the kinematical properties of the globular clusters in the Milky Way galaxy. Posti $\&$ Helmi (2019) followed up an earlier work by Binney $\&$ Wong (2017), and described the population of the Milky Way globular clusters by the two-component distribution function that describes a flat, rotating disk  and a spherical extended halo. Compared to the paper of  Binney $\&$ Wong (2017), Posti $\&$ Helmi (2019) used larger data set, and varied the mass of the dark matter halo and its shape. Vasiliev (2019) published the proper motion catalog of 151 Milky Way globular clusters with estimated uncertainties smaller than 0.1 mas yr$^{-1}$ for most objects. Catalog of Vasiliev (2019) complements the measurements provided by the Gaia collaboration (Helmi et al.,2018)  and HST-based determinations  Sohn et al. (2018), and is in excellent agreement with both.

Vasiliev (2019) compared his results with the ground based measurements, and found that  they are not only much less precise than the space-based ones, but often deviate from the space-based determinations by a far larger amount than implied by the quoted uncertainties. Combining proper motion measurements with the distance and and line-of-sight velocity measurements from the literature, Vasiliev (2019) found that within inner 10 kpc the population of the globular clusters has a mean rotational velocity of 50 --80 km s$^{-1}$ with nearly isotropic velocity dispersion about 100 -- 120 km s$^{-1}$.

Baumgardt et al. (2019) re-derived the mean proper motions and space velocities for 154 Milky Way globular clusters using a combination of the Gaia DR2 proper motions and ground-based line-of-sight velocities of individual member stars. Mean proper motions of Baumgardt et al. (2019) are in a good agreement with measurements of Gaia Collaboration (Helmi et al.,2018) and with Vasiliev's (2019) measurements, and also are in a good agreement with the proper motions derived by Sohn et al. (2018) from HST data if the systematic error of 0.10 mas yr$^{-1}$ in the Gaia proper motions is taken into account.

\noindent
Good quality kinematical data of nearly all known globular clusters of the Milky Way galaxy allow one to address the problem of assigning  them  to different  Milky Way components. A number of recent papers have been devoted to the problem of classification of GSc and the understanding the their origin (Myeong et al. 2019; Massari et al., 2019; Piatti, 2019; Horta et al., 2020;  Forbes, 2020; Perez-Villegas et al., 2020).

The investigation of  Massari et al. (2019) is particularly relevant here. This work essentially provides a solution to the problems of the separation of GCs into the subsystems of the Galaxy, and, most importantly, answers the question about the origin of the halo GCs. To separate the GCs between internal (in-situ) and external (ex-situ),  the authors use various dynamical properties, namely, orbital parameters like the apocenter (apo), maximum height from the disc ($Z_{max}$), and the orbital circularity parameter $\rm Circ$. The latter parameter is defined as ${\rm Circ} = L_z/L_{z}^{circ}$ where $L_z$ is the $Z$-component of the angular momentum of the cluster, and $L_{z}^{circ}$ is the angular momentum of the cluster placed on the circular orbit with the same total energy. Combination of these three parameters with the age-metallicity relation allowed to solve the problem of the separation.

In this paper we readdress the problem of defining globular clusters membership to Milky Way different components, -- the bar/bulge, the disk, and the halo only on the basis of the dynamical properties of the clusters. Basing on the clusters orbit parameters such as apo and $Z_{max}$, we first select those clusters that belong to the bar/bulge of the Milky Way and to its halo similar to how Massari et al. (2019) did. To separate the globular clusters within the thick disk region of the Milky Way galaxy into the disk and halo fractions, we use instead of the circular parameter $\rm Circ$ the distribution of the clusters on the ratio of the $Z$ projection of the angular momentum of the clusters to their eccentricity $e$, $L_z/e$, and demonstrate that this quantity allows us to distinguish clusters that belong to the halo and to the thick disk effectively.

In detail, the goal of this paper  is to study the properties of the globular cluster subsystems based on the dynamical properties of the their orbital parameters and to test the proposed criterion using accepted models of the Galactic potential.

 \noindent
The paper is organised as follows. In Section 2 we describe available kinematical data, and derive  spatial velocity components of the globular clusters in the Galacto-centric coordinate system. In this Section we also describe  the Galactic potential models (both axisymmetric and barred ones), and perform  orbit integration of the clusters. Section 3 is devoted to a new method for separating globular clusters into Galactic subsystems, an analysis of the kinematic characteristics of the resulting subsystems, and a comparison of our results with the results of Massari et al. (2019). Section 4 is devoted to testing the separation algorithm against different potential models. Section 5 discusses results of the separation in a barred potential with different bar rotational velocities. In Section 6 we add metallicity properties of the globular clusters to support our findings.  Finally, section 7 summarises our main results. The Appendix contains several tables with clusters' orbital parameters.

\section{Galactic potential model and Globular Clusters orbits}

Globular clusters are distributed over a wide range of distances. To integrate their orbits
one needs therefore a model of the Milky Way potential reliable in a wide range of distances from the Galactic center out to the very distant halo.
Bajkova $\&$ Bobylev (2016, 2017) specified the parameters of a few models of the Milky Way galaxy. They assumed that the potentials of the disk and of the bulge are described by the Miyamoto $\&$ Nagai (1975) and the Plummer models, respectively, while for the halo potential they adopted the models of Allen $\&$ Santillan (1991), Wilkinson $\&$ Evans (1999), Navarro, Frenk $\&$ White (1997), the spherical logarithmic potential of Binney (Binney $\&$ Tremaine 1987), the Plummer sphere, and the Hernquist (1990) potential. To fix the parameters of the models, Bajkova and Bobylev (2016, 2017) used observational
data, covering galactocentric distances up to 200 kpc. Within 20 kpc, Bajkova and Bobylev  additionally constrained the models by the line-of-sight velocities of the molecular clouds in their tangential points and by the kinematical data for 130 masers with known trigonometric parallaxes. At larger distances, Bajkova $\&$ Bobylev (2016, 2017) took into account the rotational velocities of the objects taken from the catalogue of Bhattacharjee et al. (2014). Adjustment of the parameters of the Galactic potential was done also by taking into account the local density of the Galactic disk $\rho_\odot=0.1 M_\odot$ \rm pc$^{-3}$ and the value of the force per unit mass perpendicular to the Galactic plane $K_{z=1.1}/2\pi G =77 M_\odot$\rm pc$^{-2}$.
Bajkova $\&$ Bobylev (2026, 2017) find that model of Navarro, Frenk $\&$ White (1997) (NFW) gives the
smallest residuals in the least square fit of the rotation curve of the Galaxy. Therefore, we use the NFW
model, modified by Bajkova $\&$ Bobylev (2016, 2017) of the Galactic potential to integrate the orbits of the globular clusters. We refer to this potential with the acronym NFWBB.

The axisymmetric gravitational potential of the Galaxy is a superposition of the potentials of the three main components: the central  spherical bulge, $\Phi_b(r(R,Z))$ the disk,  $\Phi_d(r(R,Z))$,  and the spherical halo,$\Phi_h(r(R,Z))$:
\begin{equation}
\begin{array}{lll}
\Phi(R,Z)=\Phi_b(r(R,Z))+\Phi_d(r(R,Z))+\Phi_h(r(R,Z))
\label{pot}
\end{array}
\end{equation}
with the potentials given by the expressions:
\begin{equation}
\Phi_b(r)=-\frac{M_b}{(r^2+b_b^2)^{1/2}},
\label{bulge1}
\end{equation}

\begin{equation}
\Phi_d(R,Z)=-\frac{M_d}{\Biggl[R^2+\Bigl(a_d+\sqrt{Z^2+b_d^2}\Bigr)^2\Biggr]^{1/2}}.
\label{disk}
\end{equation}

The halo potential is described by the NFW model:

\begin{equation}
\Phi_h(r)=-\frac{M_h}{r}\ln {\Biggl(1+\frac{r}{a_h}\Biggr)}.
\label{halo1}
\end{equation}
In the equations (\ref{bulge1}) -- (\ref{halo1}), $M_b$, $M_d$, $M_h$ are the masses of the subsystems, and $a$, $b$ are the parameters, determining the spatial density distributions in the bulge, in the disk, and in the halo of the Milky Way galaxy. Here, $r$ is the galactocentric radius, and $X$, $Y$, $Z$ are the Cartesian coordinates with $X$-axis directed from the Galactic center towards the Sun, $Y$-axis in the direction of the Galactic rotation, and $Z$-axis directed perpendicular to the Galactic plane toward the North Galactic Pole, $R^2=X^2+Y^2, r^2=R^2+Z^2$.

Table \ref{t:1} lists the values of the best-fit parameters of the Galactic potential model, described by the equations (\ref{pot}) -- (\ref{halo1}). We assume the distance of  the Sun to the center of Galaxy equal to $R_0=8.3$~kpc, and the velocity of the local standard of rest is equal to $V_\odot$=244 km s$^{_-1}$ in accordance with Bajkova $\&$ Bobylev (2016, 2017). The peculiar velocity of the Sun relative to the local standard of rest is taken from Sch\"onrich  et al. (2010): $(u_\odot,v_\odot,w_\odot)=(-11.1,12.2,7.3)$~km s$^{-1}$.

With the gravitational constant equal to unity ($G=1$), and with the unit of length equal to 1 kpc, the parameters of the Galactic potential model, listed in the Table \ref{t:1}, are given in units of 100 km $^2$ s$^{-2}$ for the potential, and for the mass is in units of 2.325$\times$ 10$^7$ $M_\odot$.

The axisymmetric potential was also used in the model including the
Galactic bar, where, however, the bulge mass $M_b$ is reduced
by the mass of the bar, to conserve mass.

We adopted a triaxial
ellipsoid model following Palou\v{s} et al. (1993):
\begin{equation}
  \Phi_{bar}(R,Z,\theta) = -\frac{M_{bar}}{(q_{bar}^2+X^2+[Y\cdot a_{bar}/b_{bar}]^2+[Z\cdot a_{bar}/c_{bar}]^2)^{1/2}},
\label{bar}
\end{equation}
where $M_{bar}$ is the mass of the bar, which is equal to $431\times$ M$_{gal}$ ($1.0\times10^{10}M_\odot$); $a_{bar}, b_{bar},$ and $c_{bar}$ are the three semi-axes of the bar $(a_{bar}/b_{bar}=2.381, a_{bar}/c_{bar}=3.03); $ $q_{bar}$ is the length of the bar;
$X=R\cos\vartheta$ and $Y=R\sin\vartheta$, where $\vartheta=\theta-\Omega_{bar} \cdot t-\theta_{bar}$,  $\Omega_{bar}$ is the circular velocity of the bar, $t$ is time, $\theta_{bar}$ is the bar orientation angle relative to Galactic axes $X,Y$, which is counted from the line connecting the Sun and the Galactic center (the $X$-axis) to the major axis of the bar in the direction of Galactic rotation. We adopted the estimates of the bar rotation velocity $\Omega_{bar}$=31, 41 $~km s^{-1} kpc^{-1}$  from Sanders et al. (2019),
55$~km s^{-1} kpc^{-1}$  from Antoja et al. (2014), and 70$~km s^{-1} kpc^{-1}$  from Debattista et al. (2002). On the other hand, the parameters $q_{bar}=8$~kpc, and $\theta_{bar}=45^{\circ}$ from  Bobylev $\&$ Bajkova (2016). All these bar parameters are listed in Table \ref{t:1}.

{\begin{table}[t]                            %% t~1
 \caption {Parameters of the galactic potential model}
 \label{t:1}
 	\begin{small} \begin{center}\begin{tabular}{|l|c|}\hline
		Parameter & Value\\\hline
$M_b$ (M$_{gal}$) & 443 $\pm$ 27 \\
$M_d$ (M$_{gal}$) & 2798 $\pm$ 84 \\
$M_h$ (M$_{gal}$) & 12474 $\pm$ 3289 \\
$b_b$ (kpc)       & 0.2672 $\pm$ 0.0090  \\
$a_d$ (kpc)       &   4.40 $\pm$ 0.73   \\
$b_d$ (kpc)       & 0.3084  $\pm$ 0.0050 \\
$a_h$ (kpc)       &    7.7 $\pm$ 2.1\\\hline
$M_{bar}$ (M$_{gal}$) & 431  \\
$\Omega_{bar}$,(km s$^{-1}$ kpc$^{-1}$)&31, 41, 55, 70\\
$q_{bar}$(kpc)              &  8.0  \\
$\theta_{bar}$ ($^{\circ}$) & 45 \\
$a_{bar}/b_{bar}$         & $2.381$  \\
$a_{bar}/c_{bar}$         & $3.03$  \\\hline
				\end{tabular}\end{center}
			\end{small}
	\end{table}}

Vasiliev (2019) used Gaia DR2 measurements and determined the proper motions for 151 globular clusters, containing almost all known objects. Vasiliev (2019) compared also his measurements with the existing proper motion catalogs and found a good agreement of his data with the Gaia collaboration (Helmi et al., 2018) and with the HST (Sohn et. al. 2018) measurements. In our study we use therefore proper motion measurements from Vasiliev (2019) catalog. Combining Vasiliev's (2019) proper motions of the globular clusters with the distances and the line-of-sight velocities from Harris (2010), one can calculate the spatial velocities and
the coordinates of the globular clusters in the Cartesian coordinate system centred on the Sun.

The knowledge of the current positions and of the velocities of the clusters together with the potential of the Milky Way galaxy allows us to determine cluster's orbits. {We integrate the orbits backward in time for 5 Gyr. Orbit integration is carried out by the Runge-Kutta algorithm of 4 orders.

For the sake of illustration, Figure \ref{f1} shows the orbits of three clusters, NGC~6440, NGC~6838 and NGC~5824,  that belong to the bar/bulge, disk, and to the halo of the Milky Way galaxy,  respectively. The orbits shown are calculated both in the axisymmetric and barred NFWBB potentials.

The parameters of all 151 globular clusters calculated in the NFWBB axisymmetric potential (the initial Cartesian coordinates $ X, Y$ and $Z$, the radial $\Pi$, tangential $\Theta$, vertical $W$ velocities, orbit parameters: the eccentricity $e$, the maximal orbit values $Z_{max}, R_{max}, r_{max}$, the $Z$-coordinate of the angular momentum $L_z$ and the total energy $E$) are given in Table A1 of the Appendix. In the last column, the identifier of the globular cluster membership to either a bar/bulge (B), or a thick disk (D), or a halo (H), obtained as a result of applying our subsystem separation algorithm, is given. The orbital parameter $e$ and the limits of variation of the $L_z$, as well as its mean value, obtained in a barred NFWBB potential are given in Table A2 of the Appendix.

\begin{figure}[t]
{\begin{center}
   \includegraphics[width=1.0\textwidth,angle=0]{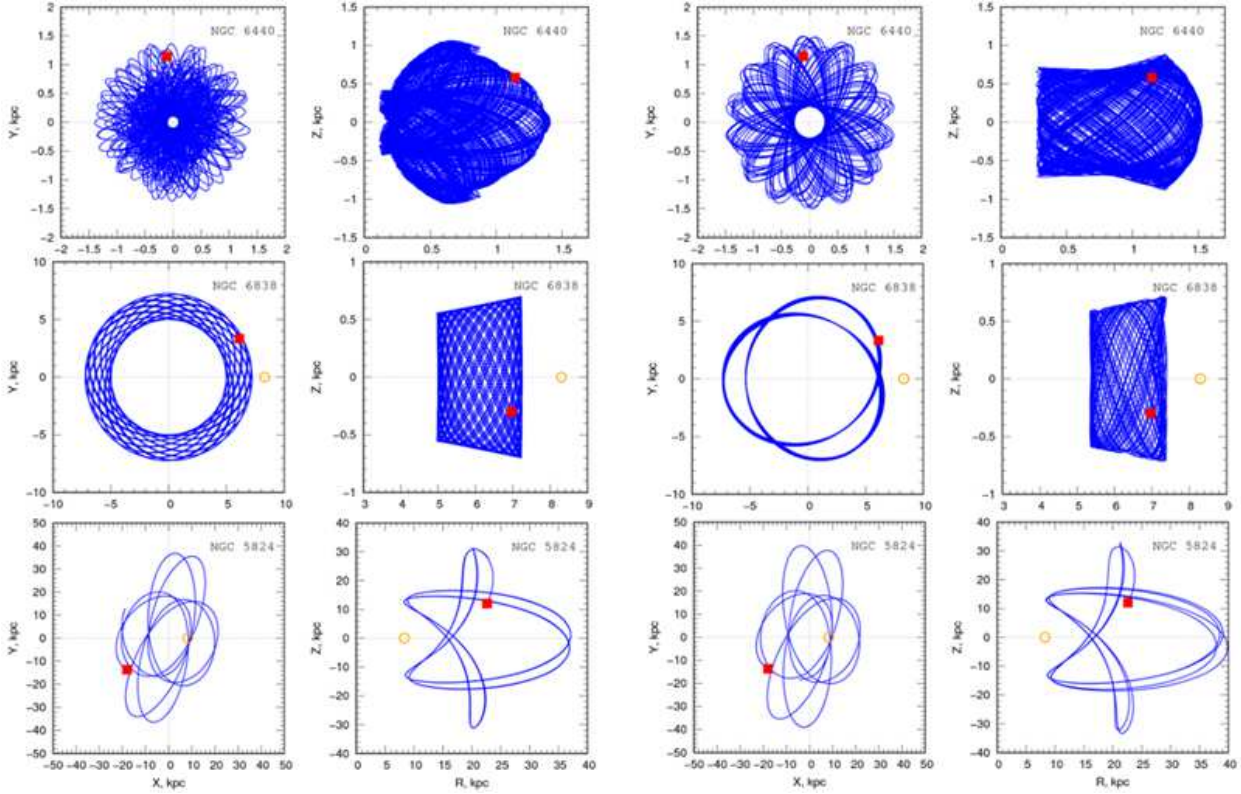}
\caption{Orbits of NGC~6440 (top line), NGC~6838 (middle line), and NGC~5824 (bottom line) belonging to  the bar/bulge, the thick disk, and the halo respectively, obtained in the axisymmetric potential (two left-hand columns) and in a barred potential with parameters from Table \ref{t:1} (two right-hand columns)}
   \label{f1}
\end{center}}
\end{figure}

\section{Criteria of separation and kinematical properties of globular clusters' subsystems}

Binney and Wong (2017) described the Milky Way globular clusters' system  using a two-component distribution function. The distribution function contains eight parameters which should be fixed using available observational data. Posti and Helmi (2018) applied this approach to a sample of 75 globular clusters from the Gaia DR2 catalog together with 20 globular clusters observed with the Hubble Space Telescope. Posti and Helmi (2018) found that halo of the Milky Way galaxy is not rotating significantly, or even has a small retrograde rotation, while the Milky Way disk  rotates with a velocity of about 210 km s$^{-1}$ at solar radius. Vasiliev (2019) used a similar approach and described the population of the globular clusters with a single distribution function. Vasiliev (2019) found that within 10 kpc from the center of the Galaxy the population of  globular clusters has a mean rotational velocity of about 50 -- 80 km s$^{-1}$, and a nearly isotropic velocity dispersion of about 100 -- 120 km s$^{-1}$. We use a different approach to assign globular clusters to the various components of the Milky Way galaxy. Namely,
we follow the idea of Posti et al.(2018) who used stars' orbits  to determine their membership to the Milky Way subsystems.

\subsection{Our separation approach}

Our approach is based on the orbital parameters  of the clusters. Intuitively, clusters with elongated orbits, or  having  orbits highly inclined to the plane of the Galactic disk belong to the Milky Way halo, while  clusters with nearly circular orbits and high circular velocities that are close to the plane are members of the Galactic disk.

The determination of globular clusters as members of  the bar/bulge, the disk, and the halo was done  in two steps. 

\noindent
Firstly, clusters probably belonging to the bar/bulge, and to the halo subsystems of the Galaxy were selected.
To this aim, we used the empirical stellar map of the bar/bulge region of the Milky Way galaxy taken from Wegg $\&$ Gerhard (2013). They measured the three-dimensional stellar density distribution of the galactic bar/bulge region covering an area of $|X|\times |Y|\times |Z|=2.2 \times 1.3 \times 1.1$ kpc  in  the inner region of the Milky Way. This density map shows a highly elongated bar with the projected axis ratios of (1 : 2.1) for isophotes reaching $\sim$ 2 kpc along the major axis. Globular clusters belonging  to the bar/bulge region would need to be searched for among those located inside a volume of ($\pm$ 2.2) $\times$ ($\pm$ 1.3) $\times$ ($\pm$ 1.1) kpc. To refine the definition, we then add the criterion  $r_{max}<3.5$ kpc.
The halo globular clusters were selected using as a criterion the height above the Galactic plane  $|Z|>6$ kpc,  assuming that the thick disk/halo separation occurs at $|Z|$ less than $6$ kpc. The resulting sample of  bar/bulge objects consists of 20 globular clusters (namely, NGC 6266, 6316, 6355, 6440, 6522, 6528, 6558, 6624, 6642, Terzan 2, 4, 1, 5, 6, 9, BH 229, Liller1, Djorg 2, ESO 456-78, Pismis 26).
The number of  globular clusters that  belong to the halo is instead equal to 56 and the number of  clusters that are left out from these criteria and then would require further separation is equal to 75.

\noindent
Secondly,  the sorting of the remaining globular clusters was done using the probabilistic approach. The approach is based on the determination the kinematical parameter of a globular cluster sample that clearly demonstrates a net bimodality in their distribution within the thick disk of the Milky Way galaxy consisting of the two different classes of objects. Namely, globular clusters belonging to the thick disk have  quasi-circular orbits with relatively small eccentricities and  large circular velocities, while  globular clusters that belong to the halo subsystem have more elongated orbits with lower circular velocities.

To split the  two classes of the globular clusters within the thick disk of the Milky Way galaxy, we use the ratio of the angular momentum, $L_z$ to the eccentricity  $e$.  The distribution  of clusters' parameter $L_z/e=\Theta\times R/e$ shows a clear  bimodality. Figure \ref{f2} illustrates the approach. Left-hand panel of the Figure shows the distribution of the globular clusters when we adopt the axisymmetric NFWBB model of the Galactic potential.
Knowing the parameters of the Gaussian distribution functions in Figure \ref{f2} which are
listed in Table \ref{t:2}, one can determine the membership probability of a particular globular cluster to the halo or to the disk of the Milky Way. We find that globular clusters belonging to the thick disk of the Milky Way galaxy consist of 35 objects (NGC 104, 4372, 5927, 6218, 6235, 6254, 6304, 6356, 6352, 6366, 6362, 6397, 6496, 6539, 6540, 6541, 6553, 6569, 6656, 6749, 6752,
6760, 6838, 7078, Pal 1, 7, 8, 10, 11, E 3, ESO 224-8, FSR 1716, BH 184, Terzan 3, Terzan 12)
The number of the halo globular clusters within the thick disk of the Milky Way galaxy that kinematically belong to the halo is equal to 39. Combining these globular clusters with the clusters that are located in the Milky Way halo, we get in total 96 globular clusters in the Milky Way halo. We underline that a close look at individual orbits of the clusters confirms that the splitting of the clusters was done consistently.

\begin{figure}[t]
{\begin{center}
   \includegraphics[width=0.3\textwidth,angle=270]{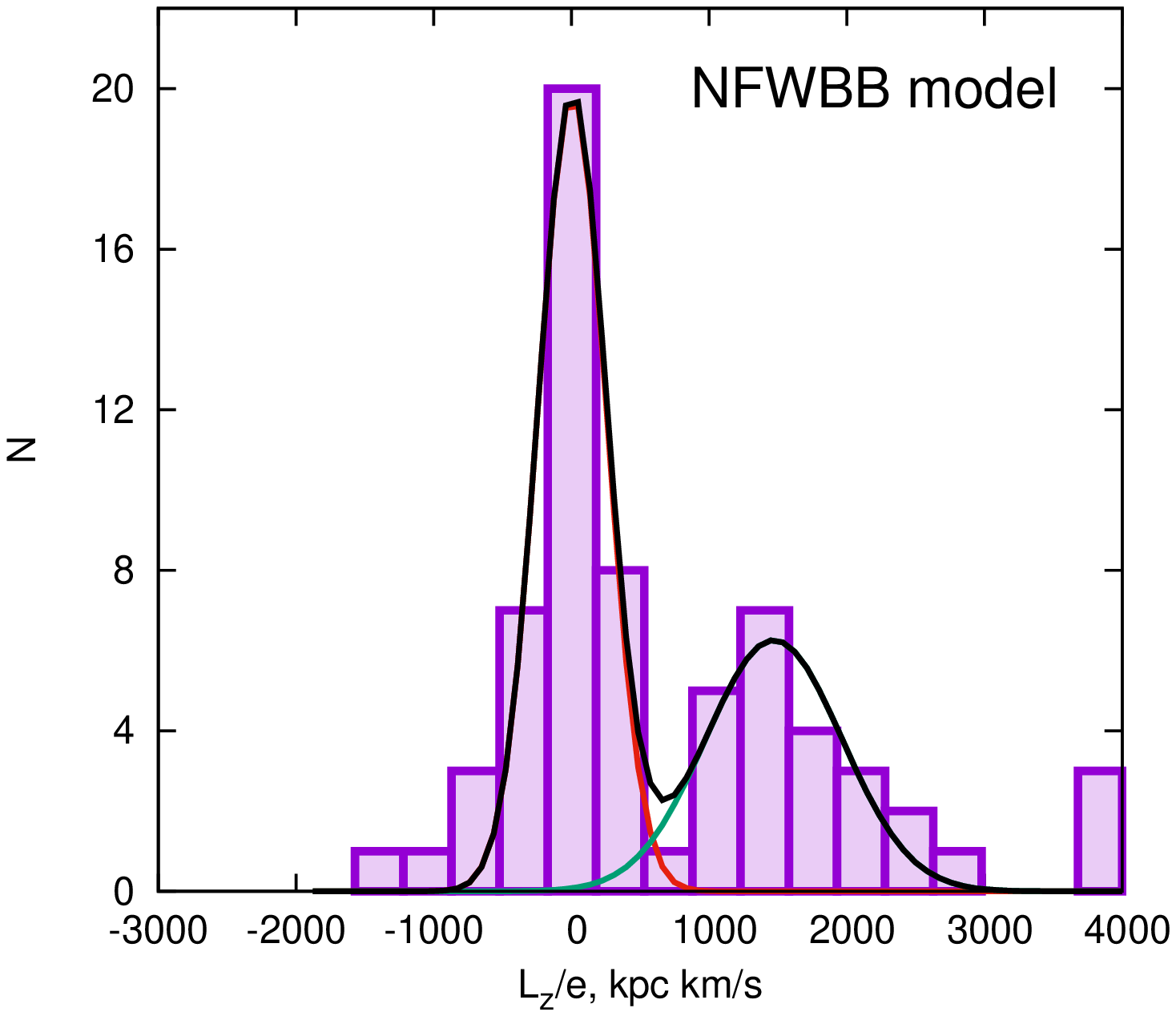}
     \includegraphics[width=0.3\textwidth,angle=270]{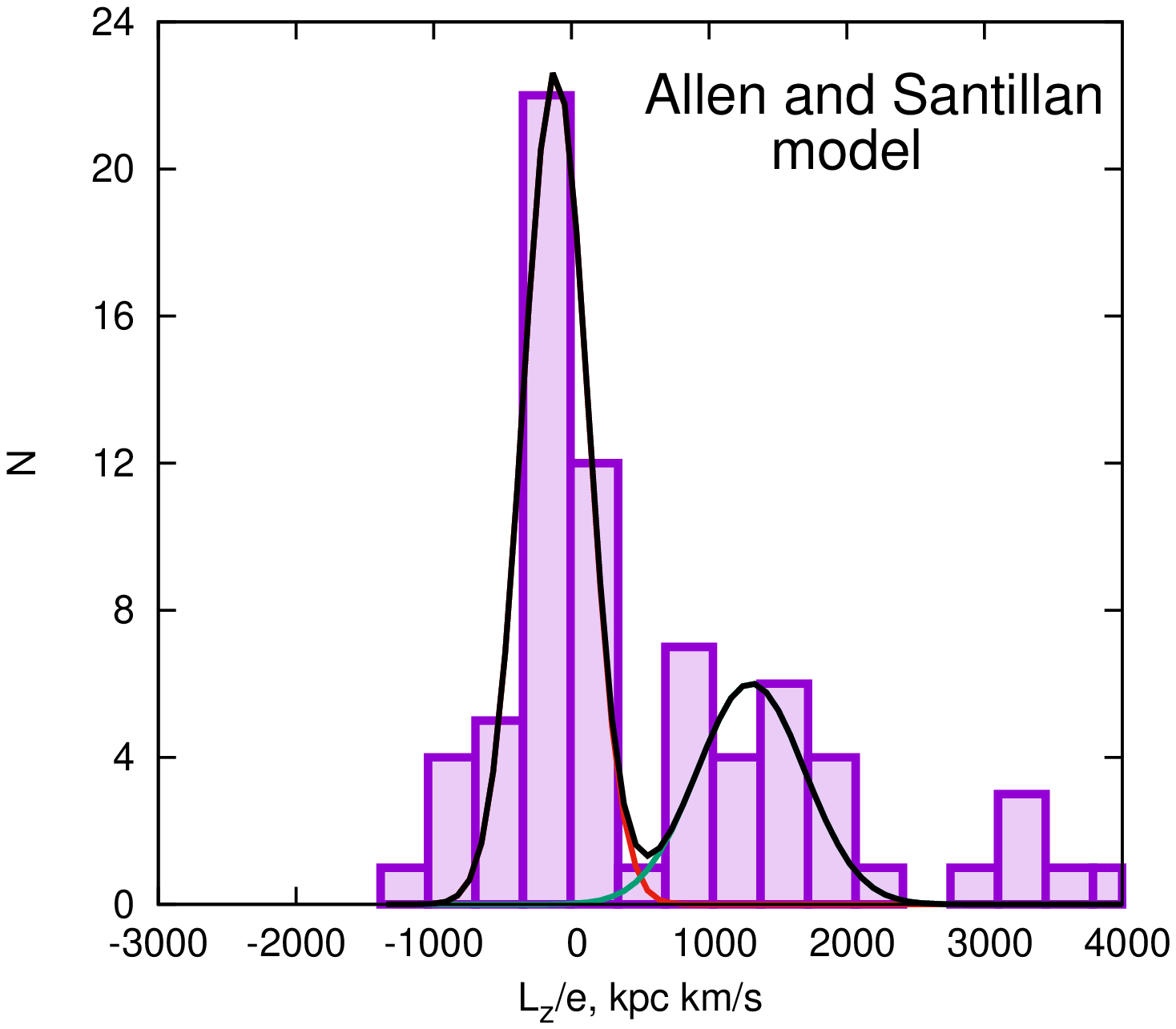}
      \includegraphics[width=0.3\textwidth,angle=270]{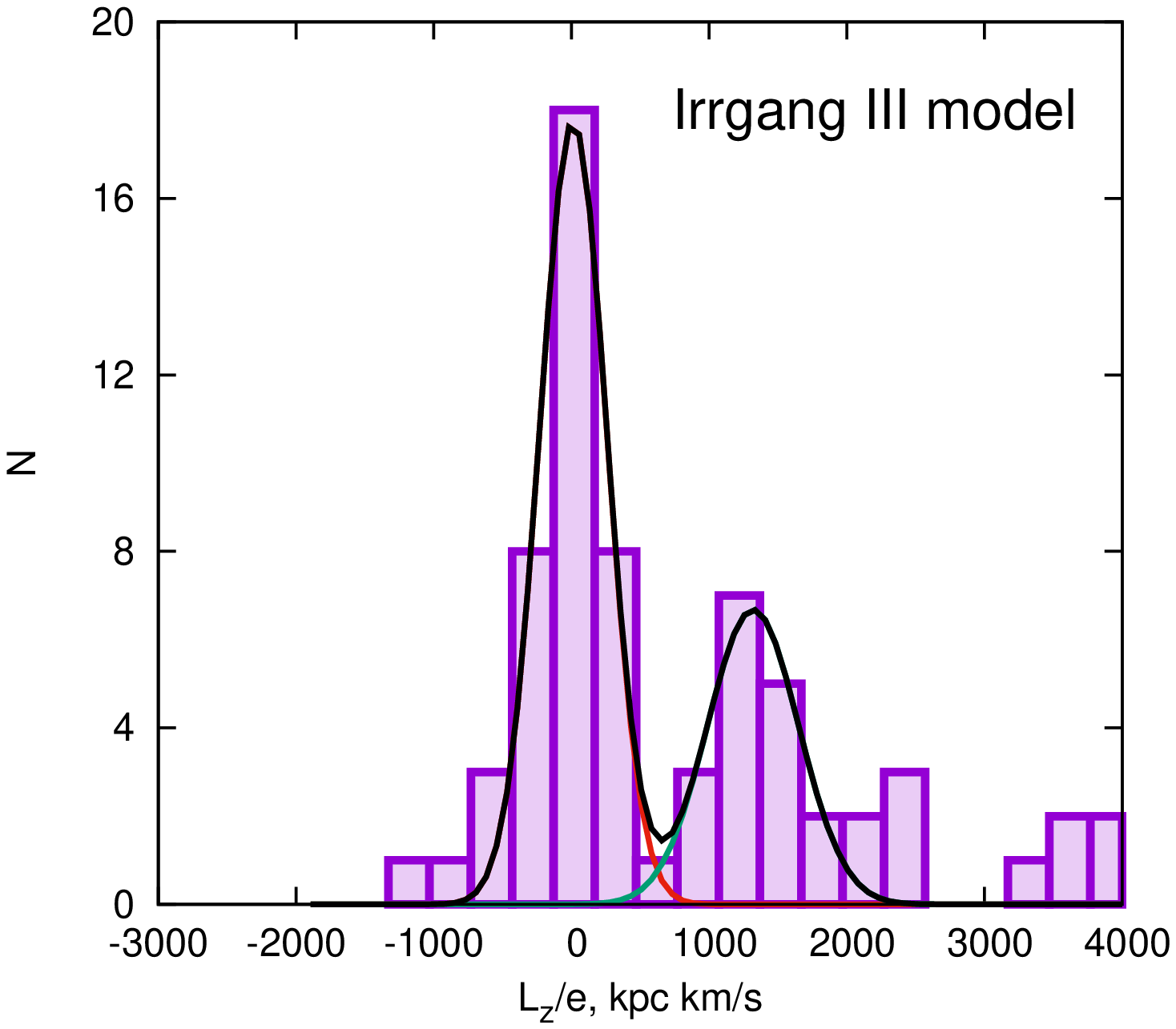}
\caption{Distribution of the globular clusters by the ratio of the angular momentum to eccentricity $L_z/e$ in the NFWBB axisymmetric model (left-hand panel), in the Allen and Santillan model (middle panel), and the Irrgang III model (right-hand panel)}
\label{f2}
\end{center}}
\end{figure}

%%%%%%%%%%%%%%%%%%%%%%%%%%%%%%%%%%%%%%%%%%%%%%%%
 {\begin{table}[t]                            %% t~2.
 \caption {Parameters of the Gaussian components in Figure \ref{f2}}
 \label{t:2}
 \begin{small}
 \begin{center}
 \begin{tabular}{|l|cc|cc|cc|}\hline
    {\small Parameter}&  \multicolumn{2}{|c|}{\small NFWBB model } & \multicolumn{2}{|c|}{\small Allen  $\&$ Santillan model }& \multicolumn{2}{|c|}{\small Irrgang III model} \\
     &  \multicolumn{2}{|c|}{\small Gaussian components:} &  \multicolumn{2}{|c|}{\small Gaussian components:} & \multicolumn{2}{|c|}{\small Gaussian components:}\\
     &{\small Left-hand}&{\small Right-hand}&{\small Left-hand}&{\small Right-hand}&{\small Left-hand} &{\small Right-hand} \\\hline
Amplitude                                      & 20 & 6  & 23 &   6  & 18 &    7  \\\hline
The mean value                  &6&1470&  -116 & 1304 &13 & 1319  \\
($m$) [kpc km s$^{-1}$] & & & & & &  \\\hline
The standard devia-   &249 &496 & 235& 390  & 244&  332  \\
tion ($\sigma$) [kpc km s$^{-1}$] & & & & & &  \\\hline
\end{tabular}
\end{center}
\end{small}
\end{table}}

The membership of individual globular clusters to  a given  Galaxy subsystem (GS) ($B$ (bar/bulge), $D$ (disk) or $H$ (halo)) is indicated in the last column of Table A1.

\subsection{Kinematical properties of Milky Way components as traced by their globular clusters}

In this Section we used the derived clusters' orbital parameters to determine the kinematical  properties of the parent Galactic component.
To this aim, we firstly characterise the uncertainties in distances, proper motions and line-of-sight velocities by employing a  Monte-Carlo approach. The artificial data sets were pulled for the bar/bulge, the disk, and the halo globular cluster samples within corresponding error bars in the distances, the proper motions, and the line-of-sight velocities. The errors in proper motions and the line-of-sight velocities were taken from  Vasiliev (2019). The distances for the artificial data sets were assigned assuming 5 and 10 percent uncertainties in the individual distances to the clusters. For each artificial data set we find the average radial  $<\Pi>$ and rotational  $<\Theta>$ velocities for the bar/bulge, the thick disk and the halo. The velocities of the subsystems of globular clusters and the determined errors are listed in Table \ref{t:3}. Figure \ref{f3} shows the histograms of the GCs velocity distributions in the bar/bulge, the thick disk, and the halo of the Milky Way galaxy. Figure \ref{fE} shows distribution of the subsystems objects in the diagram "$E$ --- $L_z$". It is interesting to note the specific location of the bar/bulge, and the thick disk objects.

\medskip

\noindent{\it Bar/bulge}

\medskip

Velocities of the subsystems of the globular clusters are given in Table 3. Figure 3 (top panels) shows the velocity distributions of theglobular clusters belonging to the bar/bulge of the Milky Way galaxy. We find the average rotational velocity of the globular clusters within the bar/bulge of 49 +/- 11 km/sec. The Milky Way bulge kinematics has been a subject of extensive studies in recent years. Using Bulge Radial Velocity Assay survey (BRAVA) Howard et al. (2008) find that the rotational velocity of bulge stars changes between -100 and ~ + 80 km/sec within  $-10^o \leq l \leq+10^o$. This estimate agrees with Kunder et al. (2012) who find that the rotational velocity of bulge stars changes from -100 to ~ 80 km/sec over the bulge region they surveyed. Ness et al. (2012) also find that for  $-10^o \leq l \leq +10^o$  the bulge rotational velocity changes within -100  ~ +80 km/sec (see their Figure 3).Finally, Zoccali et al. (2014, their Fig.~6) using the GIRAFFE inner bulge survey find that rotational velocity of
sample of bulge red clump stars changes from $\sim +75$ to  $\sim -75$ km/sec. All these studies  confine the rotation curve in
Galactic longitude $-10^o \leq l \leq +10^o$, which translates into cartesian Galacto-centric coordinate $Y$ in the range $\pm1.3$ kpc.
A close inspection at Table~A1 shows that all the globular cluster marked as bulge clusters satisfy this constraint.
With the poor statistics of our bulge globular clusters we cannot clearly derive the bulge rotation curve. Still, our estimate of the
bulge average rotational velocity as traced by globulars is
$49\pm11$ km s$^{-1}$ and falls well within the reported determinations.

\medskip

\noindent{\it Thick disk}

\medskip

Figure \ref{f3} (middle panels) shows the velocity distributions of the globular clusters belonging to the thick disk of the Milky Way. The main kinematical feature of the globular clusters within the thick Milky Way disk is their significant rotation. Our estimate of the rotational velocity of the globular clusters,
belonging to the thick Milky Way disk yields the value of 179 $\pm$ 6 km s$^{-1}$ with the lag of the rotational velocity relative to assumed velocity of LSR of 59 - 71 km s$^{-1}$. This value of the lag between the rotational velocity of the thick disk globular clusters and the rotational velocity of LSR is roughly comparable to the lag of the thick disk stars relative to the stellar Milky Way thin disk as estimated by Pasetto et al (2012) to be about 49 $\pm$ 6 km s$^{-1}$.

\medskip

\noindent{\it Halo}

\medskip

Figure \ref{f3} (bottom panels) shows the velocity distributions of the globular clusters that belong to the halo of the Milky Way. As one can clearly see from the Figure, the velocity distributions of the halo globular clusters can be approximated by a gaussian distributions both in the radial and in the azimuthal directions. The estimated values for the average velocities are: $< \Pi>= -14 \pm 4$ km s$^{-1}$, $<\Theta>= 1 \pm 4$ km s$^{-1}$. The value of the rotation velocity  is close to zero and we can conclude that the halo as traced by globular clusters does not rotate on average.

%%%%%%%%%%%%%%%%%%%%%%%%%%%%%%%%%%%%%%%%%%%%%%%%
 {\begin{table}[t]                            %% t~3.
 \caption {Average velocities of the subsystems of the globular clusters }
 \label{t:3}
 \begin{small} \begin{center}\begin{tabular}{|l|cc|cc|cc|cc|}\hline
 Galaxy        &  \multicolumn{2}{|c|}{NFWBB model (5$\%$) } & \multicolumn{2}{|c|}{NFWBB(10$\%$)} & \multicolumn{2}{|c|}{Allen $\&$ Santillan } & \multicolumn{2}{|c|}{Irrgang III } \\
subsystem      & $<\Pi>$  &$<\Theta>$ & $<\Pi>$   &$<\Theta>$ &$<\Pi>$    &$<\Theta>$ &$<\Pi>$    &$<\Theta>$\\
(N of GCs)&km s$^{-1}$&km s$^{-1}$&km s$^{-1}$&km s$^{-1}$&km s$^{-1}$&km s$^{-1}$&km s$^{-1}$&km s$^{-1}$ \\\hline

Bar/bulge (20)& 4$\pm$12&  49$\pm$11& 4$\pm$20& 50$\pm$15 &3$\pm$12 & 35$\pm$10& 4$\pm$19& 47$\pm$13\\\hline
Thick disk(35)&17$\pm$5 & 179$\pm$6 &17$\pm$7 &179$\pm$7  &15$\pm$6  &165$\pm$6 &17$\pm$6 &177$\pm$4 \\\hline
Halo (96)     &-14$\pm$4&  1$\pm$4 &-15$\pm$5 & 0$\pm$6 &-12$\pm$5 & -4$\pm$4 &-14$\pm$5& 0$\pm$5 \\\hline
\end{tabular}
\end{center}
\end{small}
\end{table}}

\begin{figure}[t]
{\begin{center}
   \includegraphics[width=0.3\textwidth,angle=270]{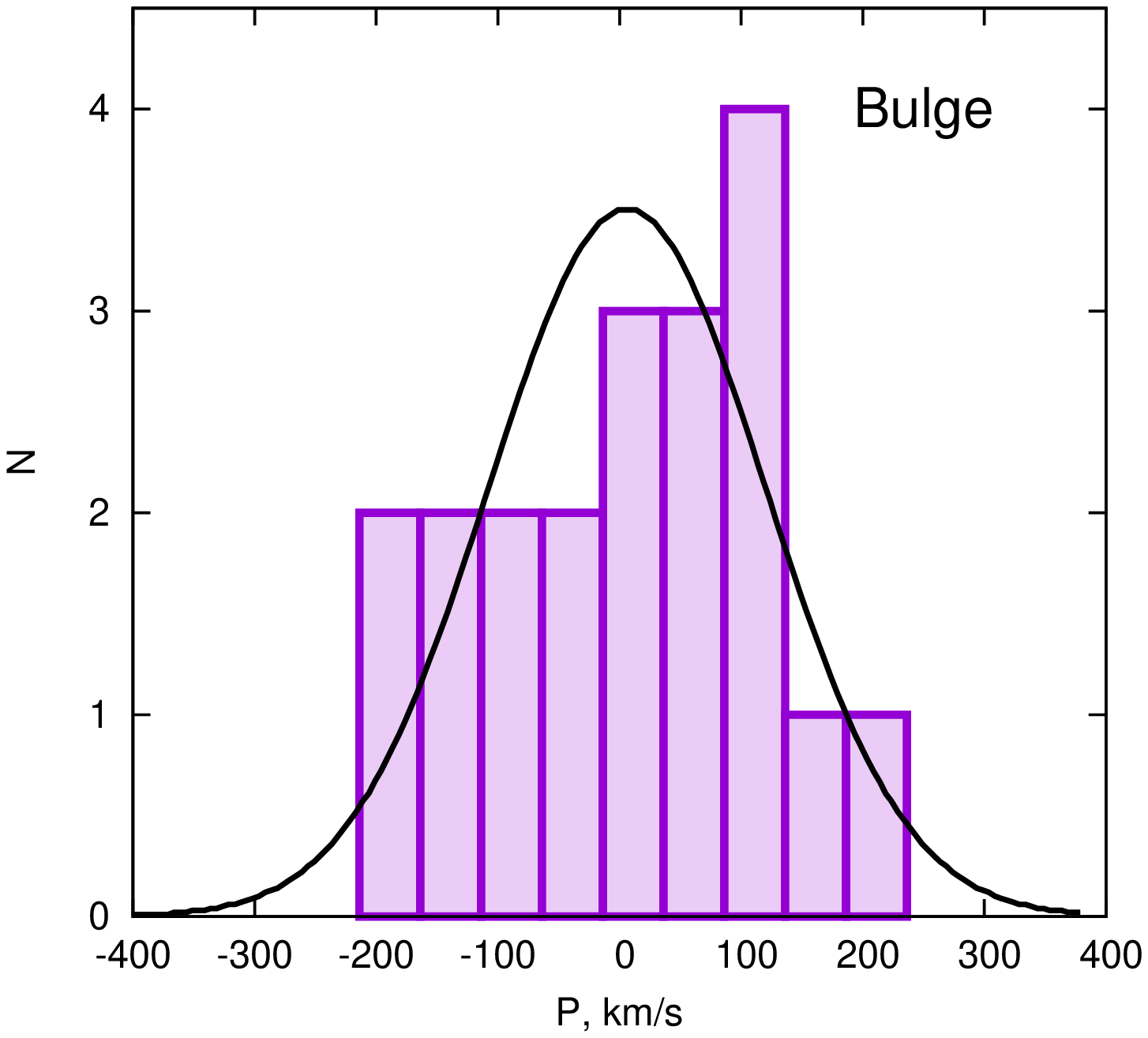}
    \includegraphics[width=0.3\textwidth,angle=270]{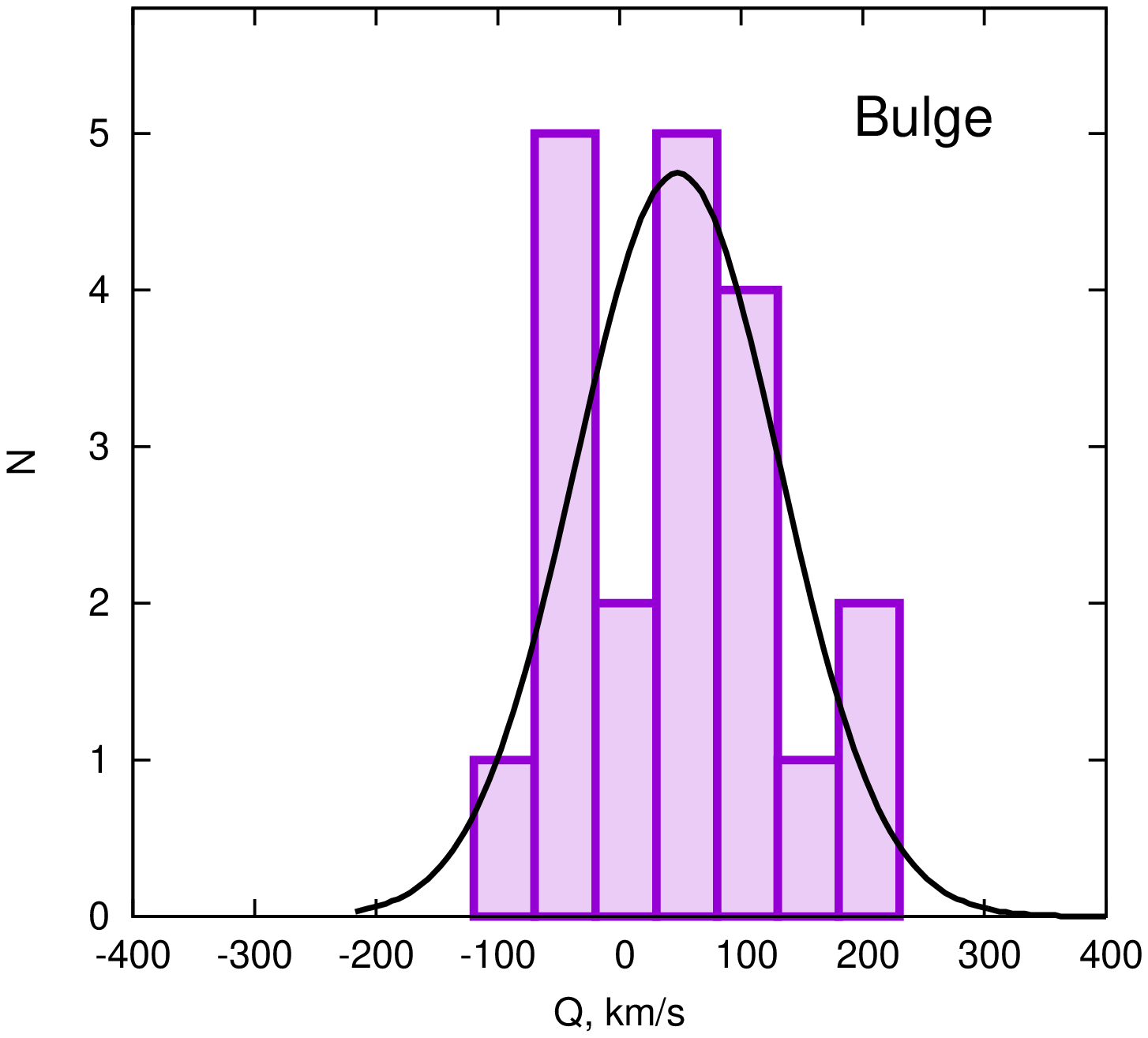}\\
     \includegraphics[width=0.3\textwidth,angle=270]{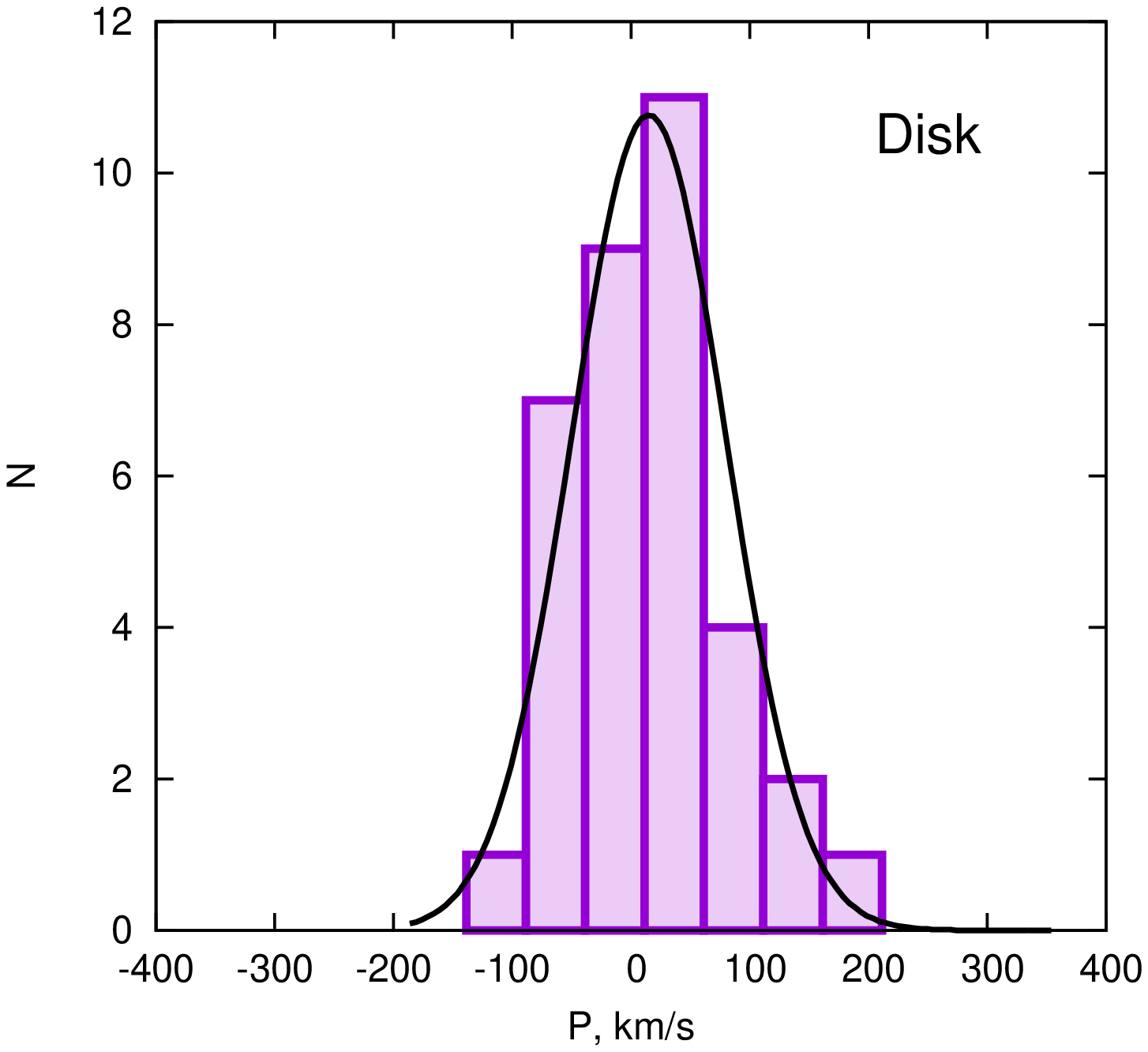}
    \includegraphics[width=0.3\textwidth,angle=270]{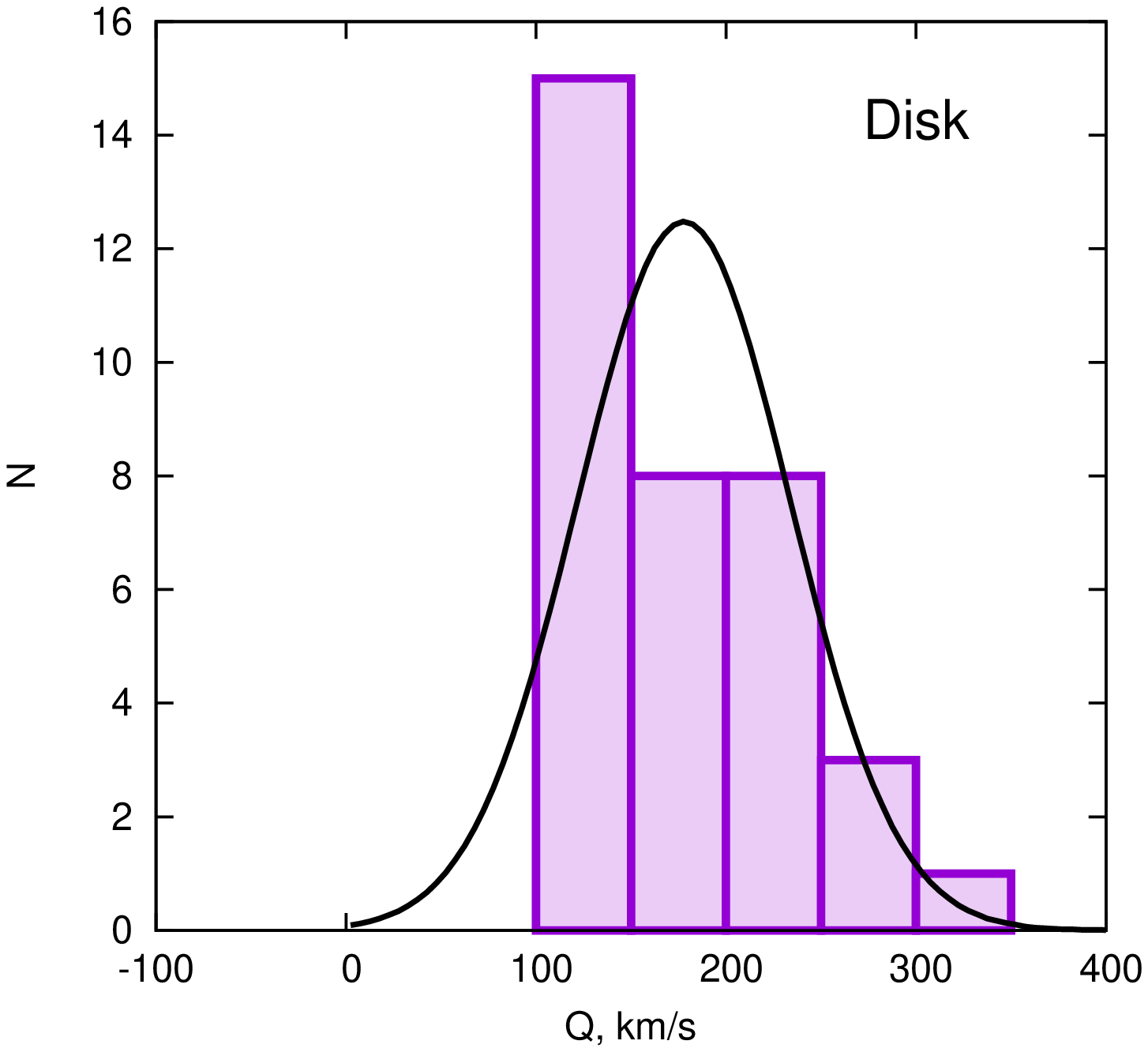}\\
    \includegraphics[width=0.3\textwidth,angle=270]{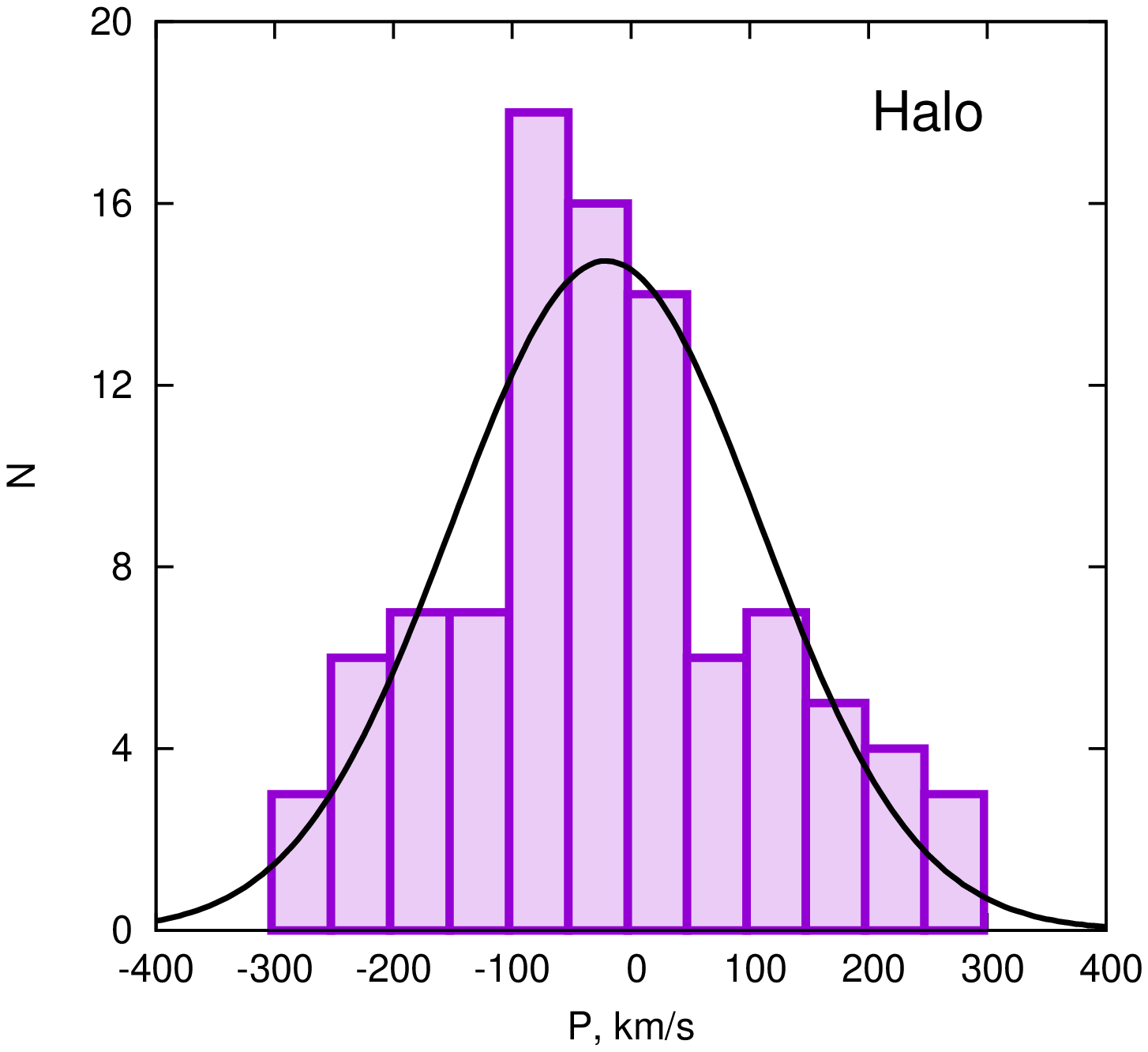}
     \includegraphics[width=0.3\textwidth,angle=270]{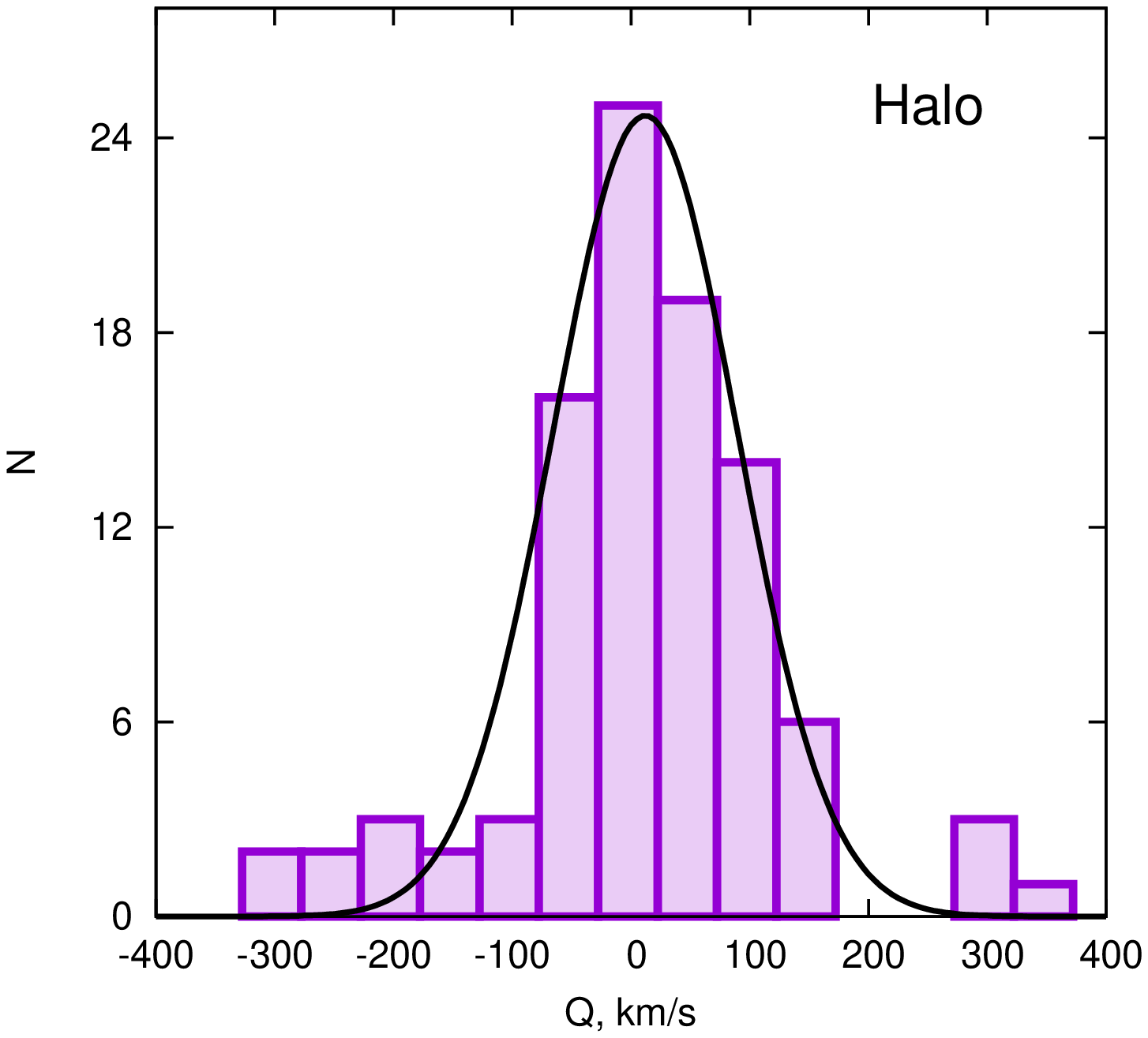}

\caption{Histograms of $<\Pi>$ and $<\Theta>$ velocity distributions for GCs of the bar/bulge (top line), the thick disk (middle line) and the halo (bottom line)}
\label{f3}
\end{center}}
\end{figure}

\begin{figure}[t]
{\begin{center}
   \includegraphics[width=0.4\textwidth,angle=-90]{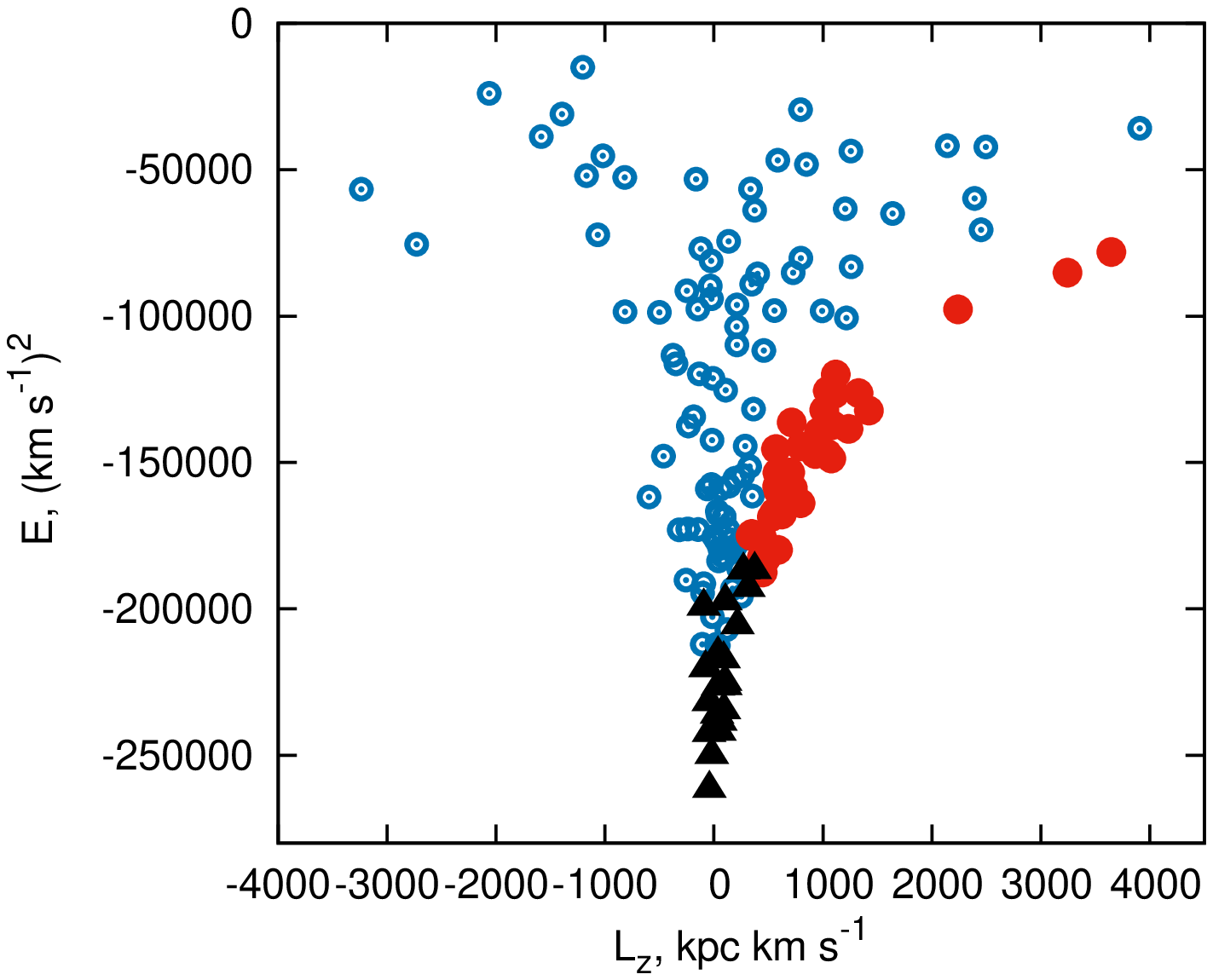}
\caption{ Distribution of the bar/bulge (black triangles), the thick disk (red closed circles), and the halo (blue open circles) GCs in the diagram "$E$ --- $L_z$"}
\label{fE}
\end{center}}
\end{figure}

\subsection{\bf Comparison with the results of Massari et al. (2019) }

We now compare our criterion based on the orbital parameter $L_z/e$ with the criterion based on the orbital parameter $\rm Circ$ (Massari et al., 2019) for separation of the GCs located in the thick disk area ($Z<5$ kpc) into the disk and  halo clusters. We restrict ourselves on the procedure described in Section 3.1 of paper by Massari et al. (2019) only,  which use purely dynamical considerations to categorise star clusters.
We stress in fact that our methodology is purely dynamical.

We made a comparison under equal conditions using the NFWBB gravitational potential and the same conditions for the preliminary selection of the bulge/bar (apo$<3.5$ kpc) and halo ($Z_{max}>5$ kpc) objects. As a result, we get 36 bulge objects and 57 halo objects.

\noindent
According to Massari et al. (2019), the separation of the remaining 58 GCs (in the band $|Z|<5$ kpc) into the disk and halo is performed using the criterion $ \rm Circ>0.5 $. Such a condition is justified by the histogram shown in their Figure \ref{f22} (left-hand panel). In this histogram the bimodality in the GCs distribution is clearly visible. Using this very same criterion, we obtained 33 candidate disk members.
The resulting list of objects differs from the list of objects obtained by Massari et al. (2019) by just five objects. These are:  NGC 6171, 6333, 6402, 6441, and Djorg 1 with high eccentricities 0.72, 0.74, 0.88, 0.66, and 0.76, respectively, and look like impostors.
Our criterion based on bimodality of GCs distribution on parameter $ L_z/e $ (see  Figure \ref{f22}, right-hand panel), returns therefore 28 disk objects which coincide with  Massari's et al. (2019) list.
Thus, using our separation criterion based on the $ L_z/e $ parameter,  we confirmed Massari et al. (2019) result which was obtained for another potential, while the use of the criterion based on the $\rm Circ$ orbit circularity parameter gave slightly different results.

\begin{figure}[t]
{\begin{center}
   \includegraphics[width=0.3\textwidth,angle=270]{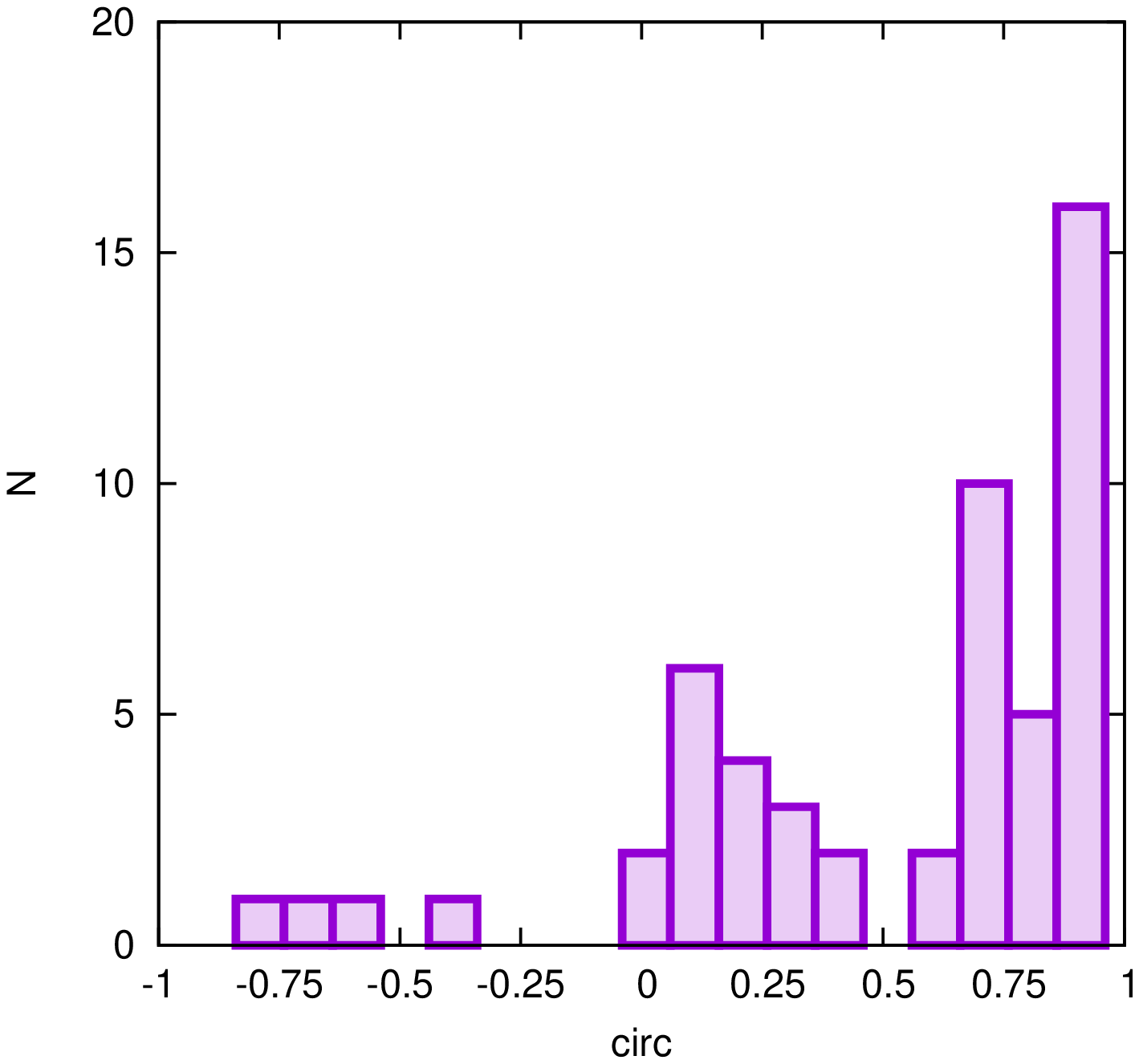}
     \includegraphics[width=0.3125\textwidth,angle=270]{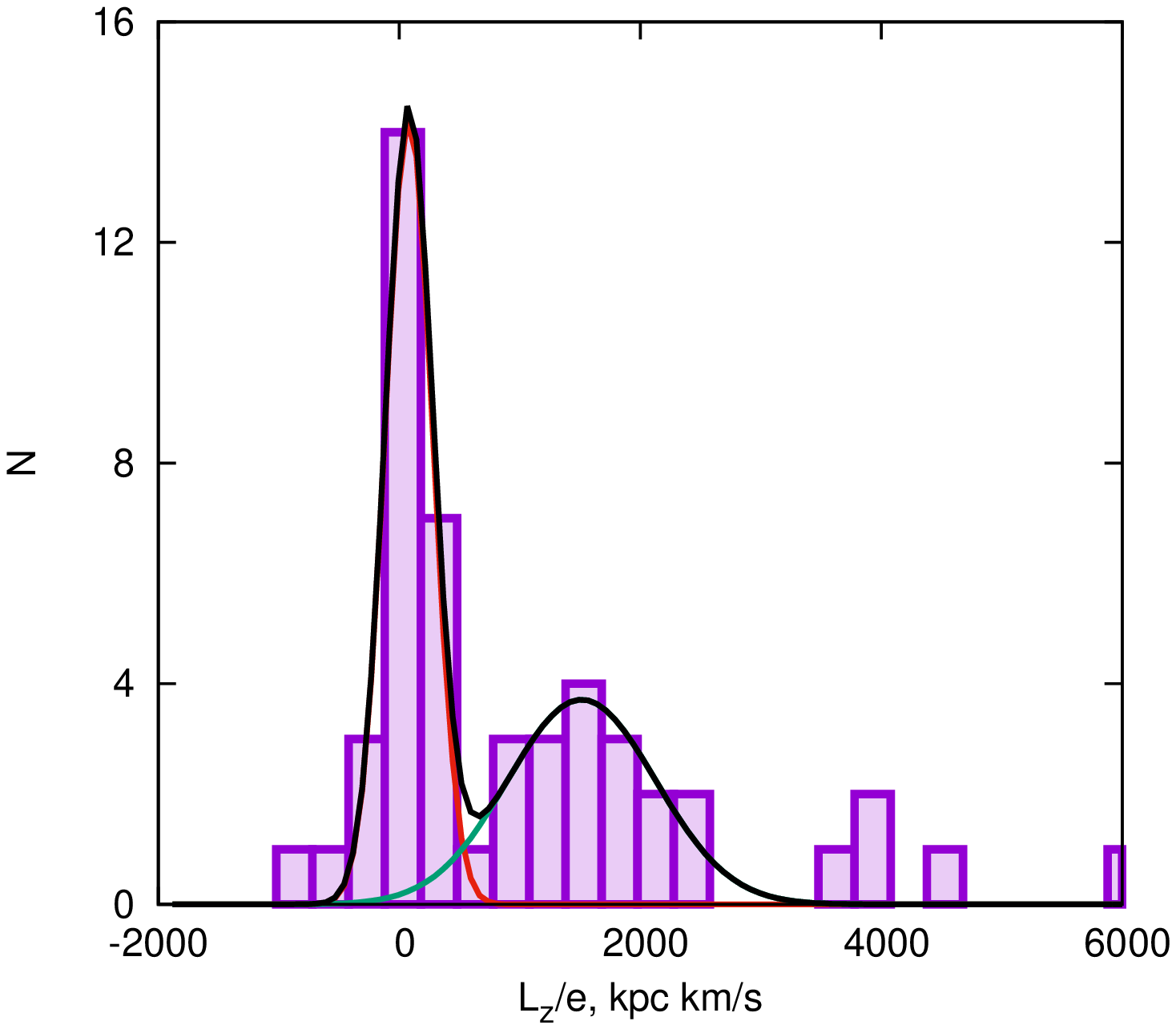}
\caption{Distribution of the globular clusters by the parameter $\rm circ$ (left-hand panel), and by the
ratio of the angular momentum to the eccentricity $L_z/e$ (right-hand panel)}
\label{f22}
\end{center}}
\end{figure}

\section{Testing the Globular Clusters separation criteria in the axisymmetric Allen $\&$ Santillan, and the axisymmetric Irrgang III potentials}

The separation criterion  was further checked against different realisations of the Galactic gravitational potential.  To this purpose, we choose the most popular models of the Galactic potential: the model proposed  by Allen $\&$ Santillan (1991), and the NFW model, modified by Irrgang et al (2013) on the basis of data on Galactic masers. We denote these potentials as the Allen $\&$ Santillan and  the Irrgang III models for the sake of brevity. The rotation curves corresponding to the NFWBB, Allen $\&$ Santillan and Irrgang III potential models are shown in Figure \ref{frot},  for comparison.

The  Allen $\&$ Santillan model assumes that Galactocentric distance of the Sun is equal to $R_\odot=8.5$ kpc, and that the circular rotational velocity of the Local Standard of Rest (LSR) is $V_\odot=220$ km s$^{-1}$, therefore the rotation curve differs considerably from the NFWBB model both at small ($R<35$ kpc) and lat arge ($R>35$ kpc) Galactocentric distances.
On the other hand, the Irrgang III model and the NFWBB model are close up to $R=20$ kpc because they were constructed on the very same masers data. This can be seen clearly from the rotation curves in Figure \ref{frot}. Both models have the same LSR velocity $V_\odot=244$ km s$^{-1}$ and  the same Galactocentric distance of the Sun $R_\odot=8.3$. At larger distances, though, the rotation curve corresponding to the Irrgang III potential goes significantly higher than ours. Such a difference in the rotation curves is due to the fact that our potential was built using data up to 200 kpc, while  Irrgang III potential employs data up to 20 kpc only.

If we apply our separation algorithm to the disk layer( $|Z|<6$ kpc) we obtain essentially the same result as employing our potential (see previous sections), despite the fact that slightly different  $L_z, e,$ and $L_z/e$ are found. This happens because, especially in the case of the Irrgang III potential, distant objects are heavily influenced by
rotation curves which deviates at large $R$. These distant objects are anyway cut off as halo objects already at the first stage of the separation algorithm. So they do not affect the separation of the GCs inside the disk layer ($|Z|<6$ kpc).
The disk objects are in fact situated within $R<20$ kpc, and the Irrgang III potential affects their selection in the same way as in the case of our NFWBB potential due to the proximity of the rotation curves  for this Galactocentric distance range. In the case of the Allen $\&$ Santillan potential, the orbits of objects within $R<20$ kpc undergo a more significant change than in the case of the Irrgang III potential. Nevertheless, the separation of objects in the disk layer $|Z|<6$ kpc remains the same due to the fact that despite the change in the parameters $L_z$ and $e$, simply because the distribution of $L_z/e$} does not change.

The comparison of orbital parameters, namely, the eccentricity $e$, the $Z$ projection of angular momentum  $L_z$, maximal values of coordinates $R_{max}$, $Z_{max}$, obtained in these three axisymmetric potentials is given in Figure \ref{fcomp1}. We stress the overall good correlation of the parameters for majority of the GCs. In particular, good correlation of parameters $R_{max}$, $Z_{max}$ is maintained  up to 20 kpc where the majority of the GCs are concentrated. At larger distances,  the mass of the halo plays an important  role. But, as we already underlined above, distant objects are cut off as halo objects already at the first stage of the separation algorithm.

The second stage of separation, based on  the bimodality of the $L_z/e$ distribution, is shown in Figure \ref{f2}, left-hand panel for the NFWBB model, middle panel for the Allen $\&$ Santillan model, and right-hand panel for the Irrgang III model. The parameters of the Gaussian components are given in Table \ref{t:2}. In spite of the differences in the distribution of the globular clusters in all these three models seen in Figure \ref{f2}, the comparison of the kinematical properties of the globular clusters in the Milky Way subsystems shows that within the errors there is no difference in the average velocities of the thick disk and of the halo in these three models (see Table \ref{t:3}). Kinematical properties of bar/bulge globular clusters are close for the NFWBB and the Irrgang III models, and differ, however, considerably from the Allen $\&$ Santillan model. Disagreement is caused, most probably, by the  difference of the disk rotational curves in its central regions in the rotation curves: the NFW models (NFWBB and Irrgang III) of the galactic potential shows a prominent hump on the rotation curve while the Allen $\&$ Santillan model simply does not show it. As one can see from Table \ref{t:3}, within the errors different globular populations clusters have the same kinematical properties in all three models of the Milky Way potential.

\begin{figure}[t]
{\begin{center}
   \includegraphics[width=0.3\textwidth,angle=-90]{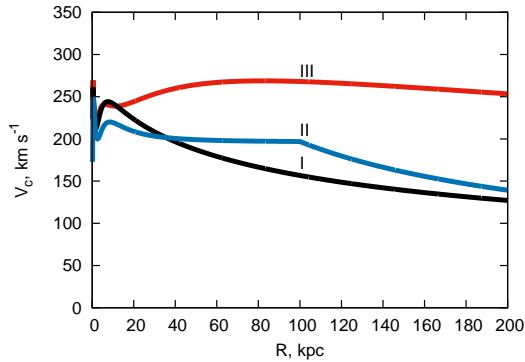}
\caption{ Rotation curves corresponding to the NFWBB (I), the Allen $\&$ Santillan (II), and the Irrgang (III) potential models }
   \label{frot}
\end{center}}
\end{figure}

\begin{figure}[t]
{\begin{center}
   \includegraphics[width=0.4\textwidth,angle=270]{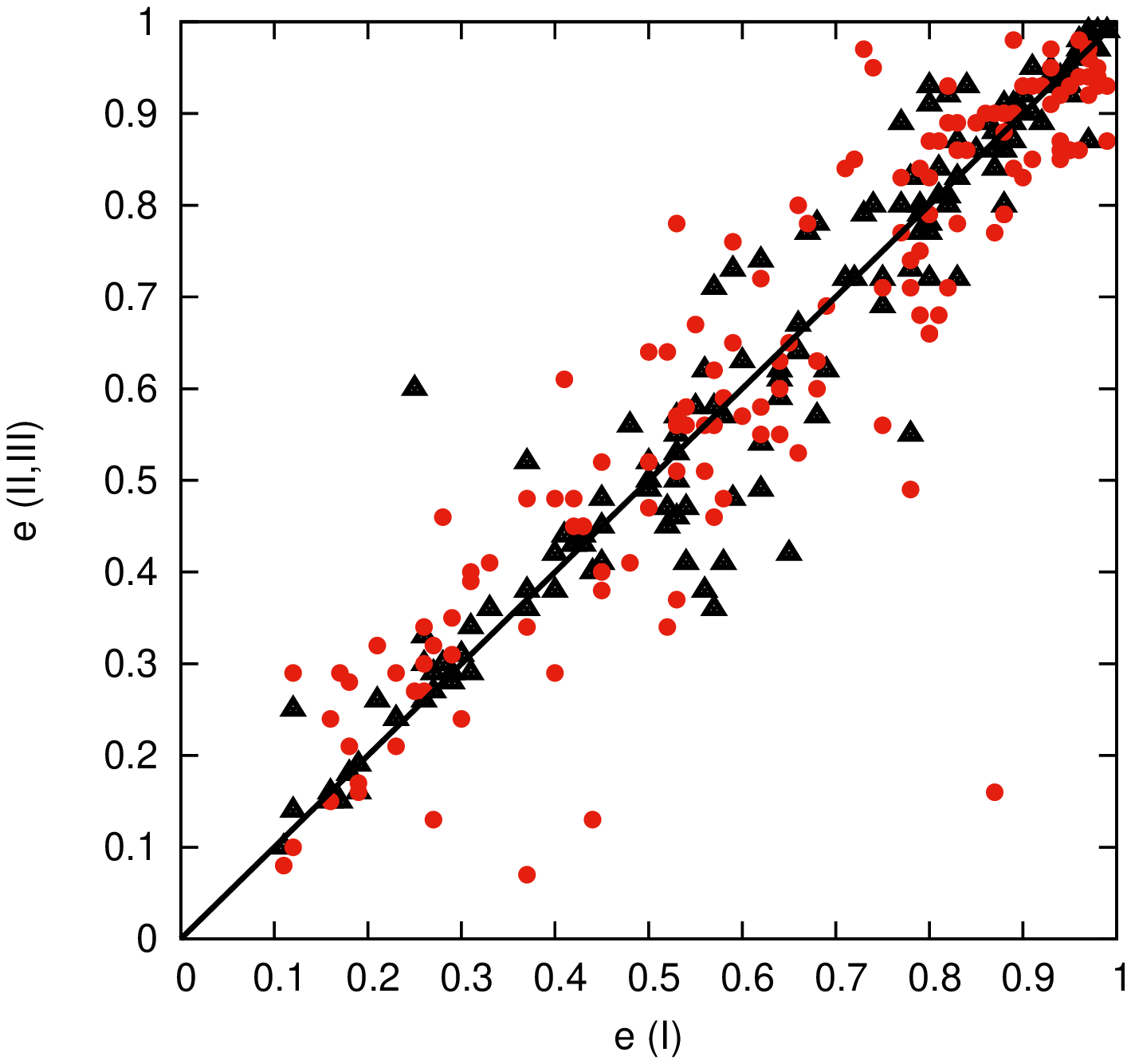}
     \includegraphics[width=0.4\textwidth,angle=270]{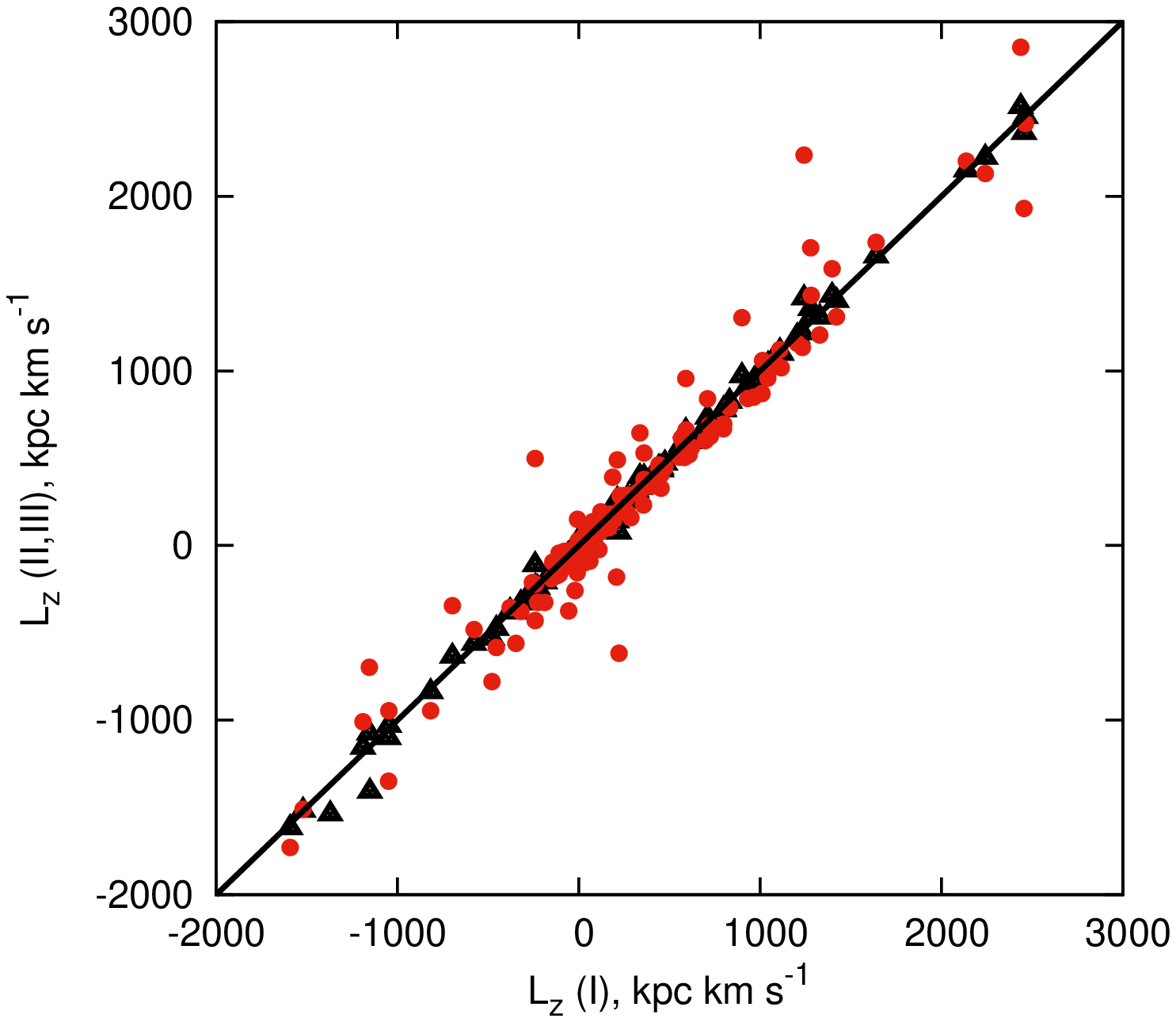}\\
      \includegraphics[width=0.4\textwidth,angle=270]{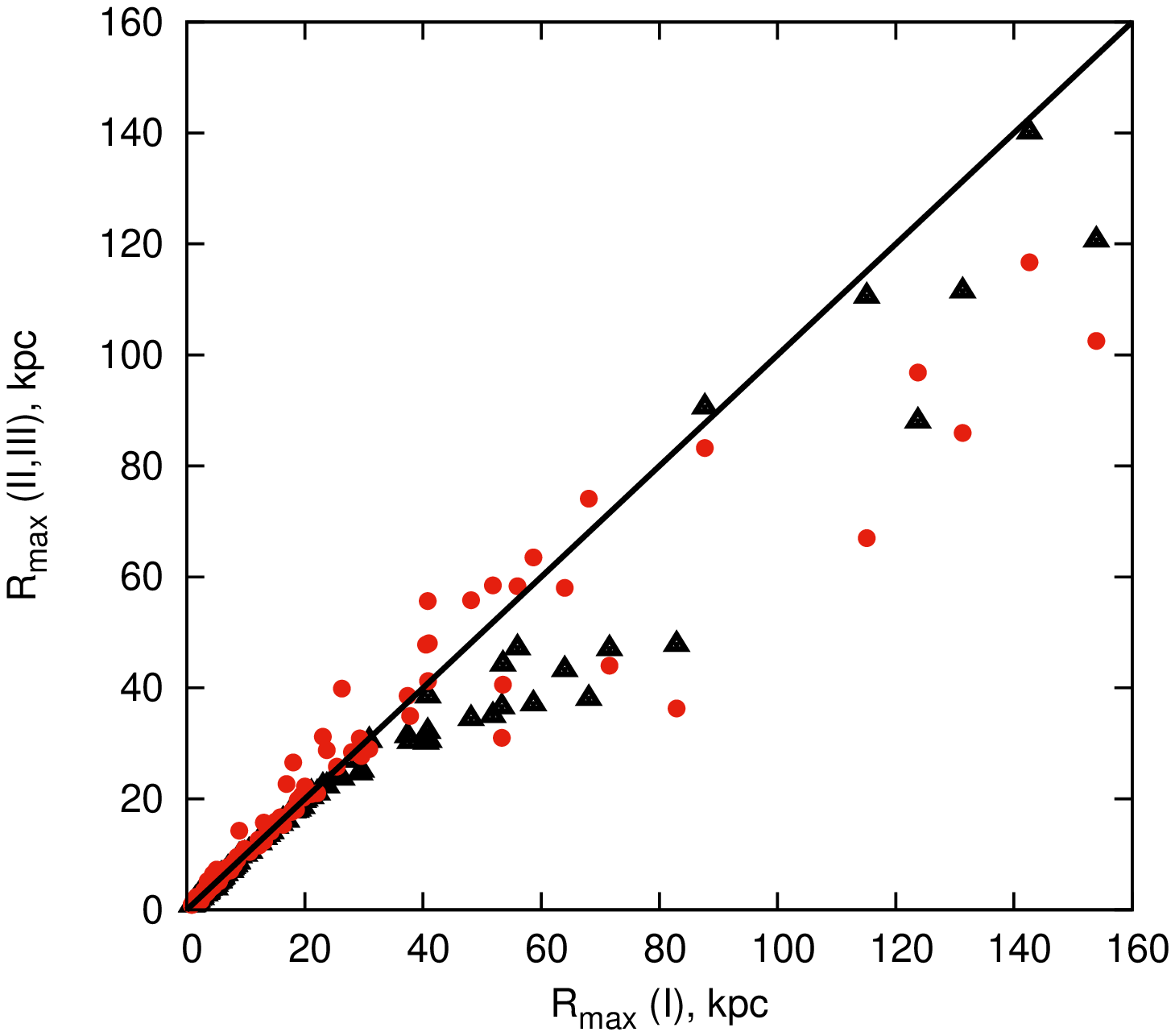}
      \includegraphics[width=0.4\textwidth,angle=270]{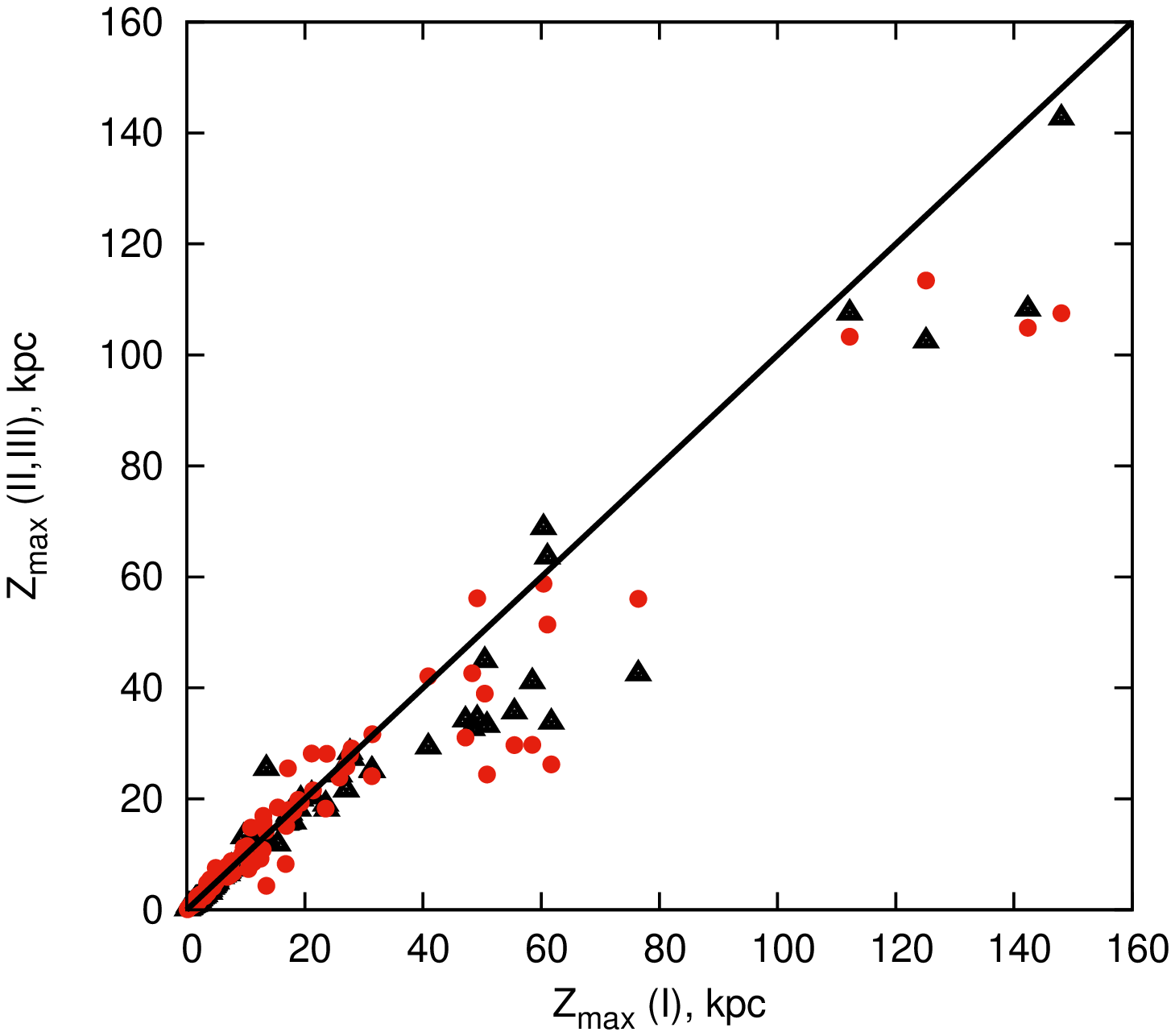}

\caption{ Comparison of the orbital parameters ($e$, $L_z$, $R_{max}$, $Z_{max}$) obtained in different axisymmetric potentials. The parameters obtained in  the Allen $\&$ Santillan (red dots) and the Irrgang III (black triangles) potentials are compared with corresponding orbit parameters obtained in the NFWBB potential. In each panel we plot the line of coincidence.}
\label{fcomp1}
\end{center}}
\end{figure}

\section{Testing globular clusters separation criteria in barred potentials}
In an attempt to better quantify the possible role of the bar/bulge potential, in this Section we test the separation algorithm in case of the NFWBB potential complemented with
the bar potential in the form of the triaxial
ellipsoid model of  Palou\v{s} et al. (1993) (see also Section 2). We adopted as mass of bar $M_{bar}=1.0\times10^{10}M_\odot$, wherein the bulge mass $M_b$ is reduced
by the mass of the bar. Since in literature the rotational velocity of the Galactic bar is reported with large uncertainties,  we checked our algorithm against four different values: $\Omega_{bar}=31, 41, 55$, and $70$ kpc km s$^{-1}$ (see Section~2). 

\begin{figure}[t]
{\begin{center}
   \includegraphics[width=0.29\textwidth,angle=270]{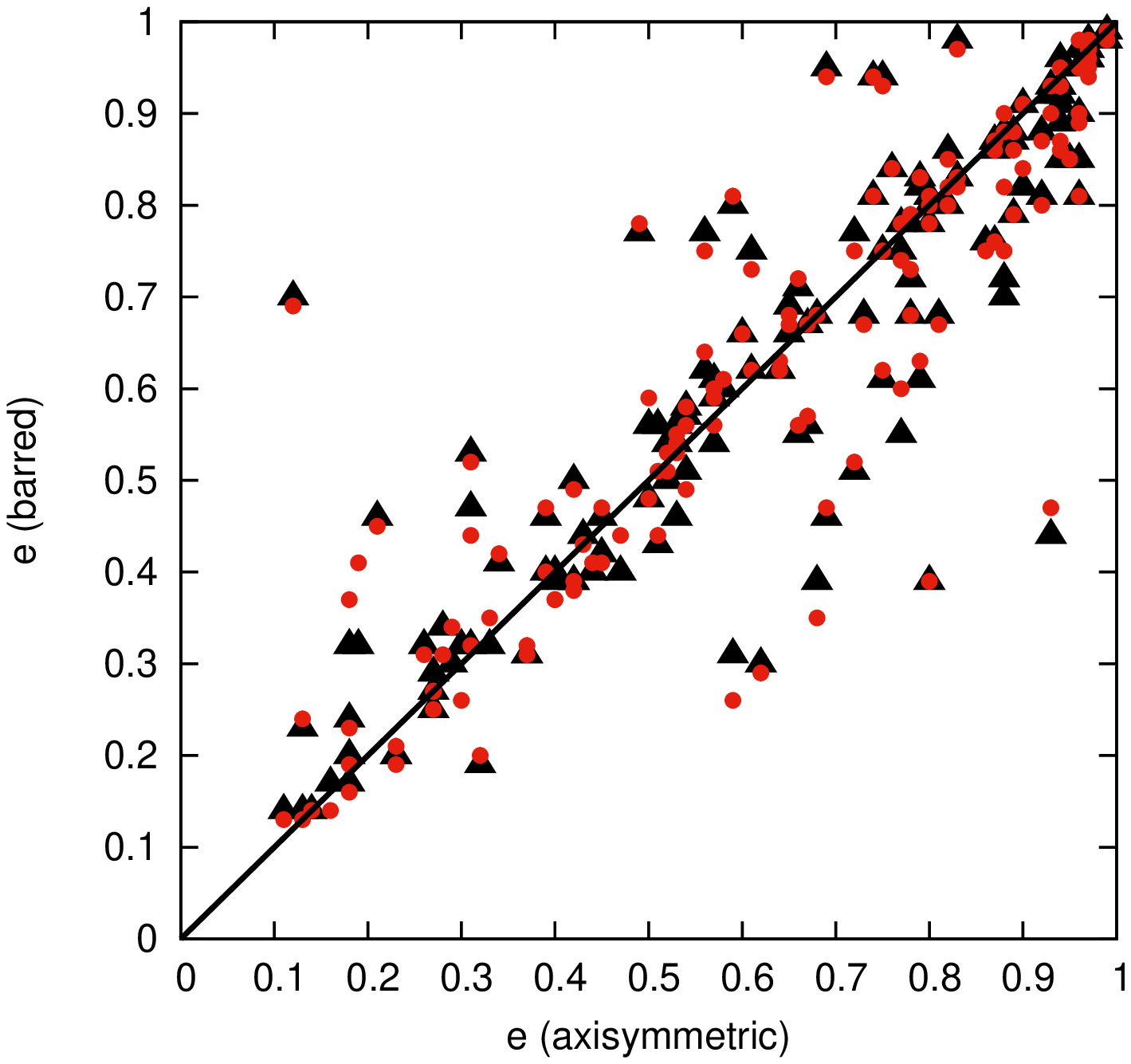}
   \includegraphics[width=0.3\textwidth,angle=270]{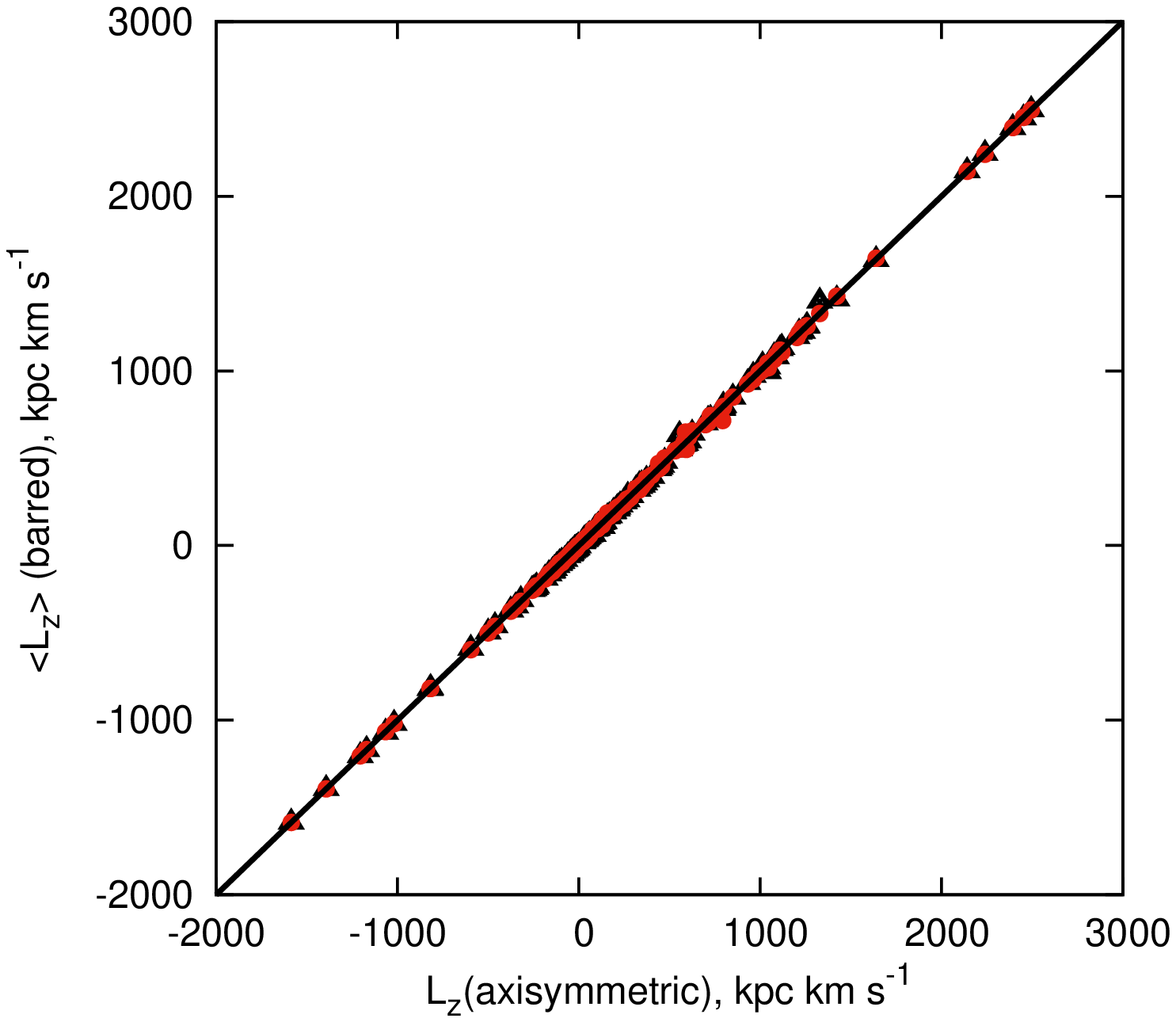}
     \includegraphics[width=0.3\textwidth,angle=270]{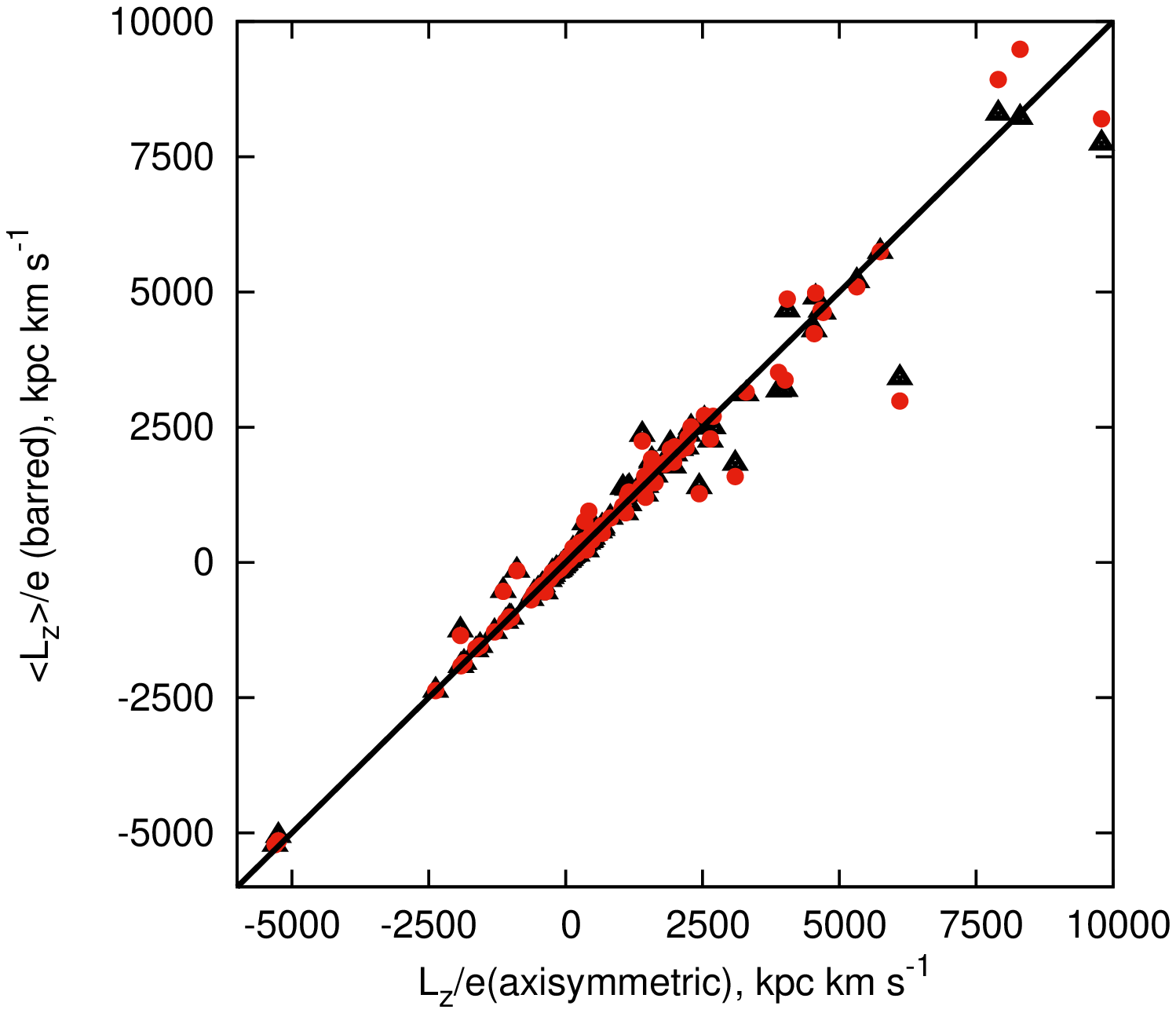}\\

     \vskip 23mm

           \includegraphics[width=0.35\textwidth,angle=0]{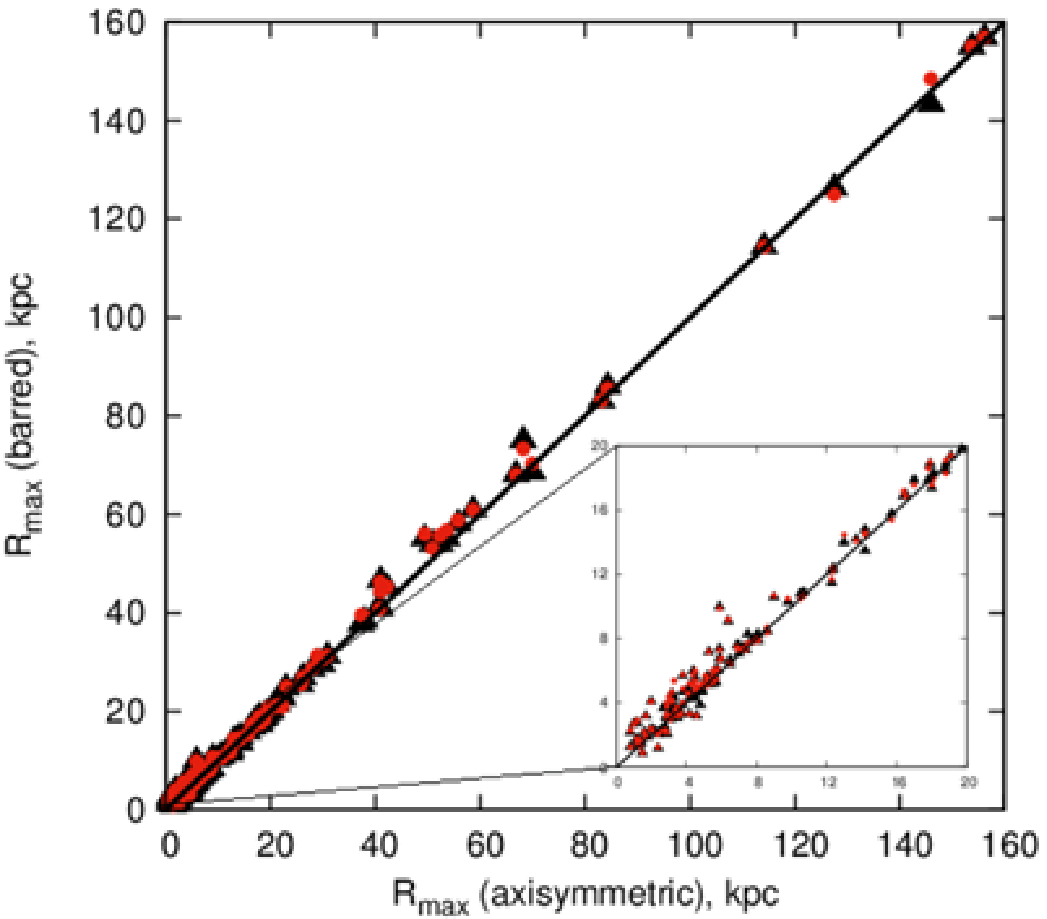} \hskip 10mm
          \includegraphics[width=0.35\textwidth,angle=0]{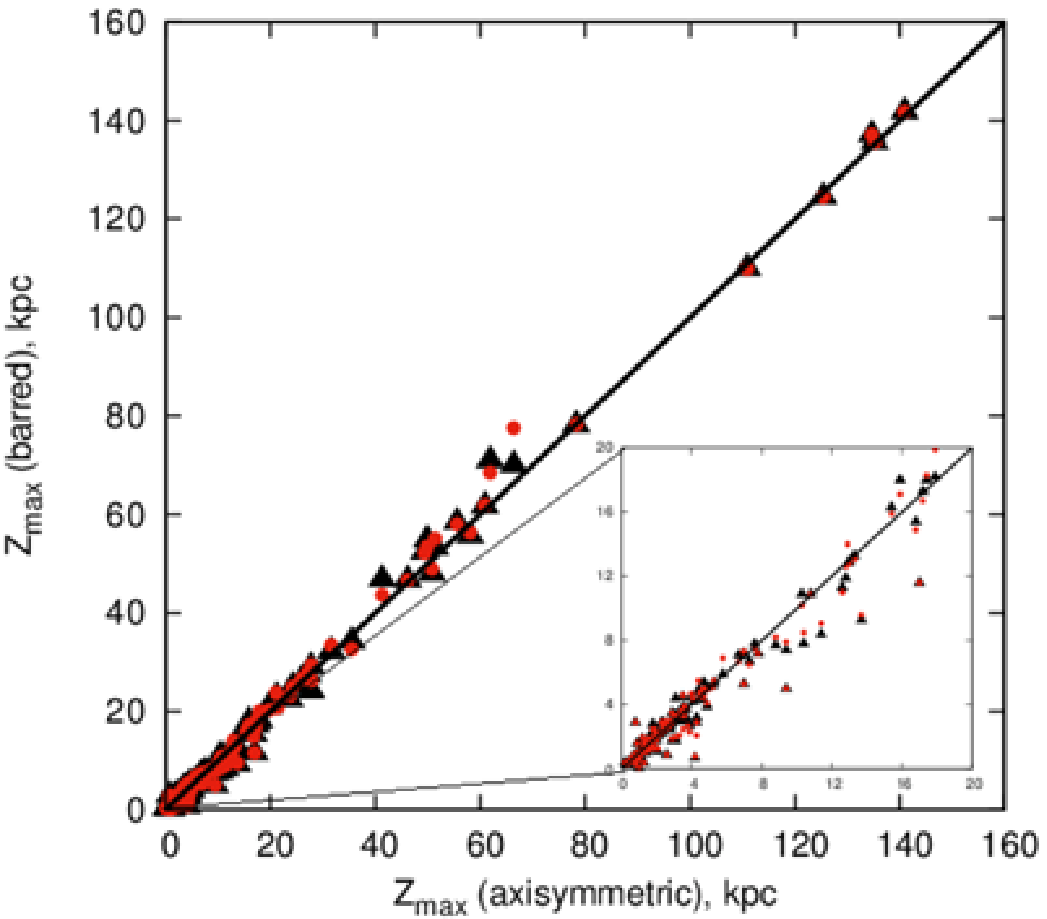}
\caption{ Comparison of the orbit parameters ($e$, $<L_z>$, $<L_z>/e$, $R_{max}$, $Z_{max}$). The parameters obtained in barred NFWBB potential with $\Omega_{bar}=55$ km s$^{-1}$ kpc$^{-1}$ (red dots), and  $\Omega_{bar}=31$ km s$^{-1}$ kpc$^{-1}$ (black triangles) are compared with corresponding orbit parameters obtained in the axisymmetric NFWBB potential. In each panel we plot the line of coincidence.}
\label{fcomp2}
\end{center}}
\end{figure}

\noindent
Our splitting algorithm stands on the  analysis of the $L_z/e$ distribution. In case of non-axisymmetric potential the $Z$ projection of angular momentum  $L_z$ is not constant along the orbit and therefore we take the value $<L_z>$ equal to average value of $L_z$ along the orbit. Orbits of the GCs, especially close to the Galactic center, can be noticeably affected by the bar. Particularly, eccentricities $e$ of  GCs orbits can change significantly, thereby significantly changing the distribution $<L_z>/e$.
The orbital parameter $e$ and the limits of variation of the $L_z$, as well as its mean value $<L_z>$ for the GCs in the disk layer $|Z|<6$ kpc obtained in the barred NFWBB potential for all adopted bar rotational velocities, are given in Table A2 of the Appendix.
The comparison of the orbital parameters ($e, L_z, L_z/e, R_{max}, Z_{max}$) is given in Figure \ref{fcomp2}. The parameters obtained in barred NFWBB potential with $\Omega_{bar} = 31$ km s$^{-1}$ kpc$^{-1}$, and $\Omega_{bar} = 55$ km s$^{-1}$ kpc$^{-1}$ are compared with the corresponding orbital parameters obtained in the axisymmetric NFWBB potential.
We notice a general good correlation of the parameters for the majority of the GCs.

\begin{figure}[t]
{\begin{center}
   \includegraphics[width=0.3\textwidth,angle=270]{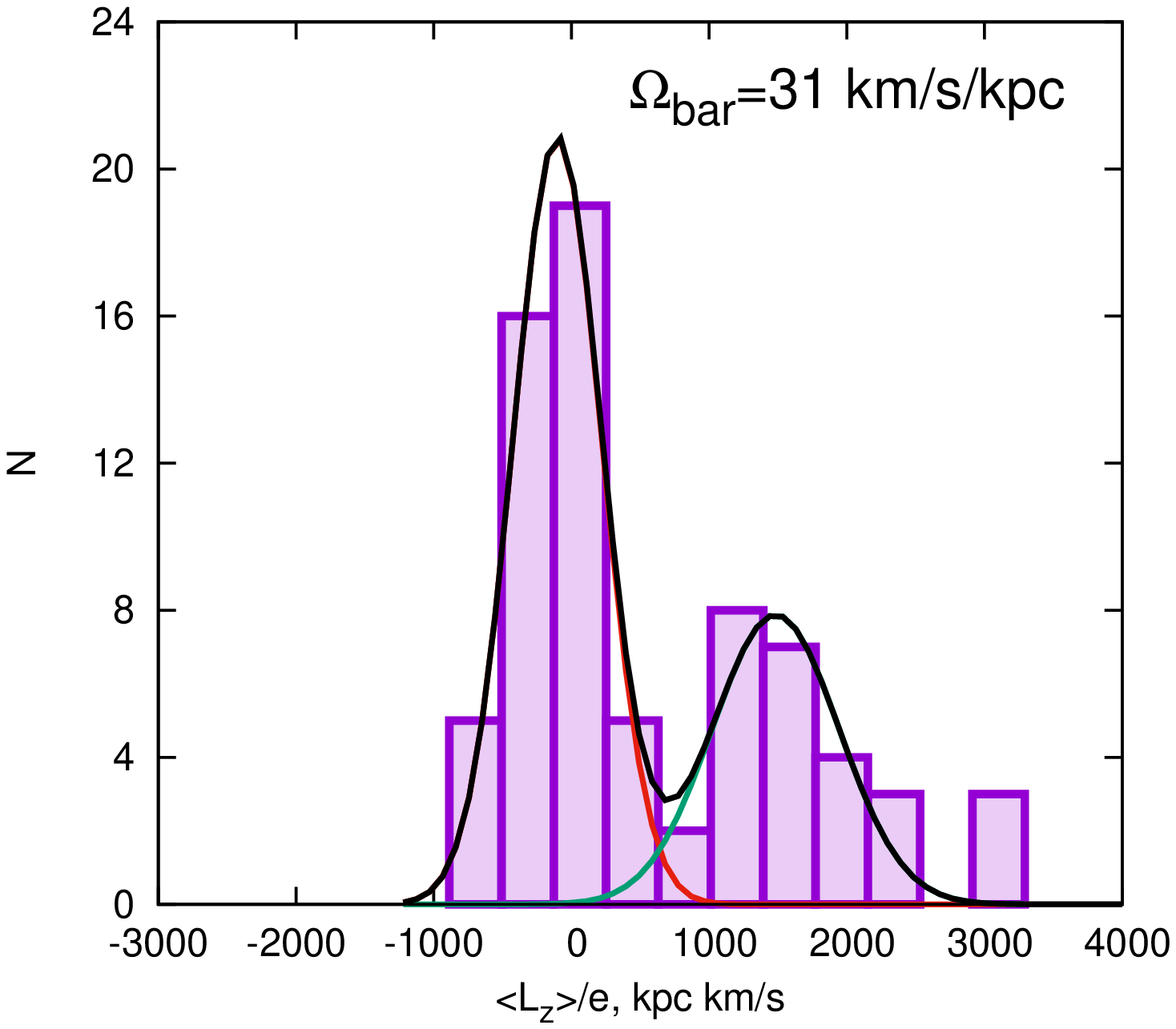}
     \includegraphics[width=0.3\textwidth,angle=270]{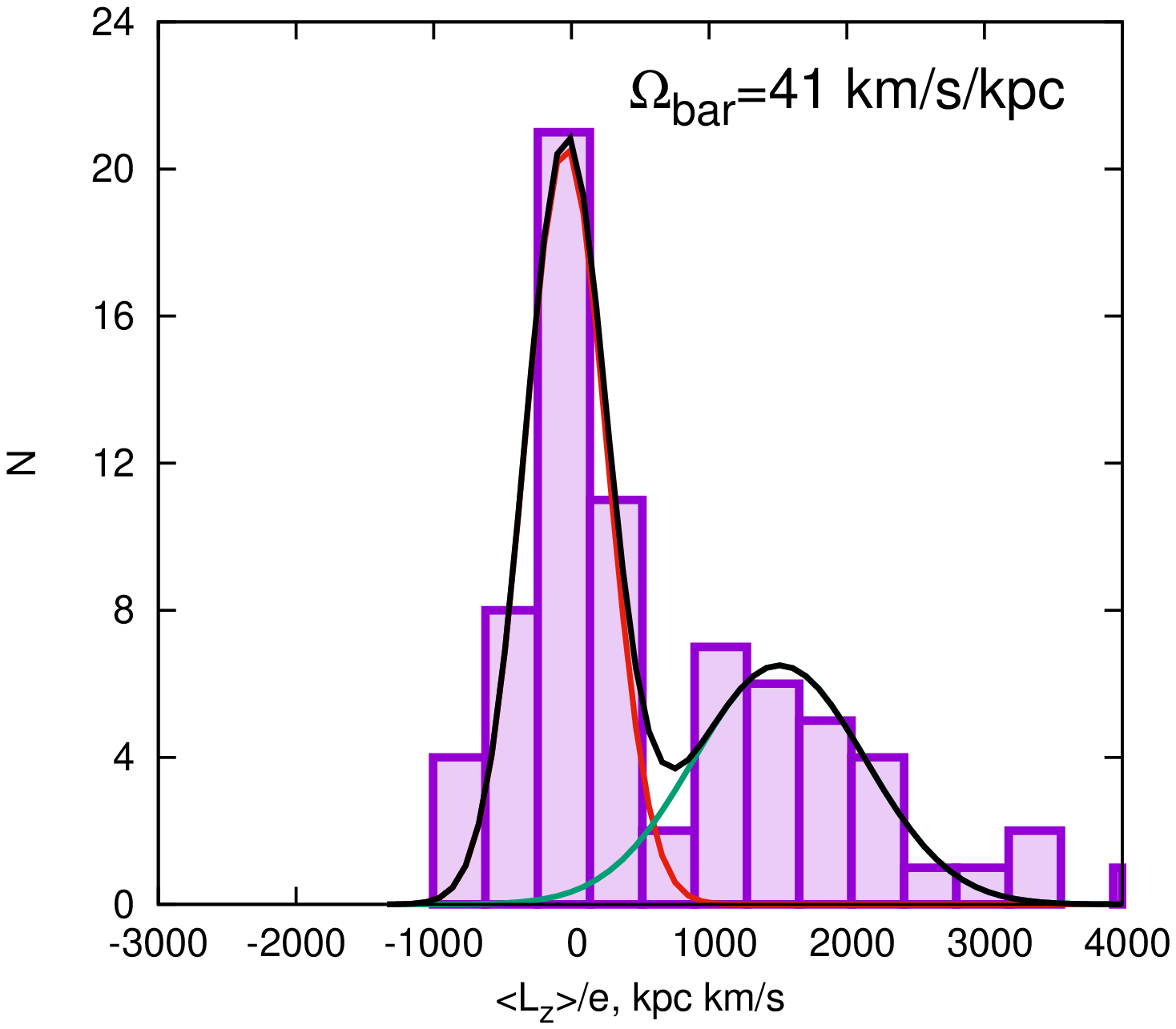}\\
      \includegraphics[width=0.3\textwidth,angle=270]{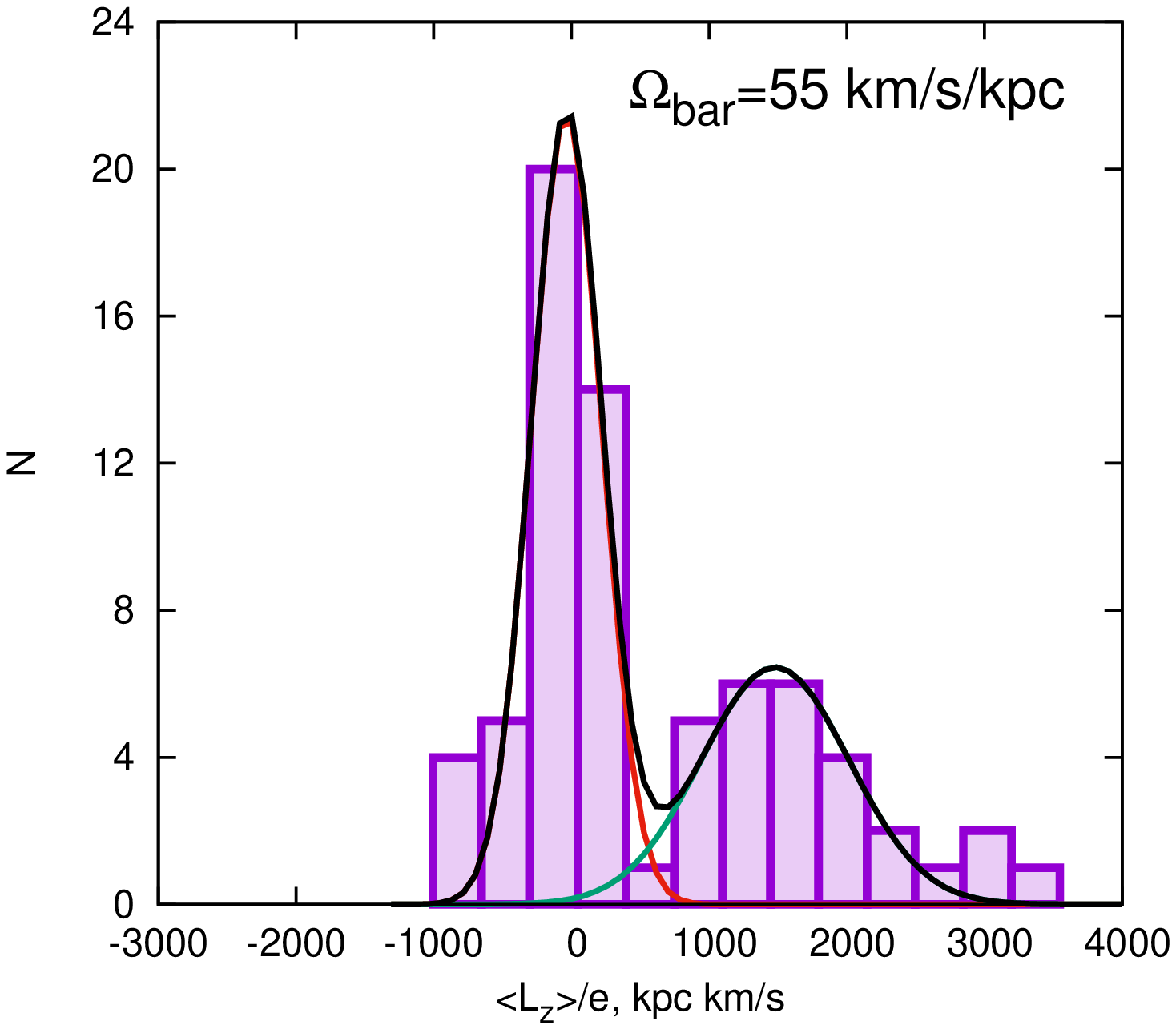}
      \includegraphics[width=0.3\textwidth,angle=270]{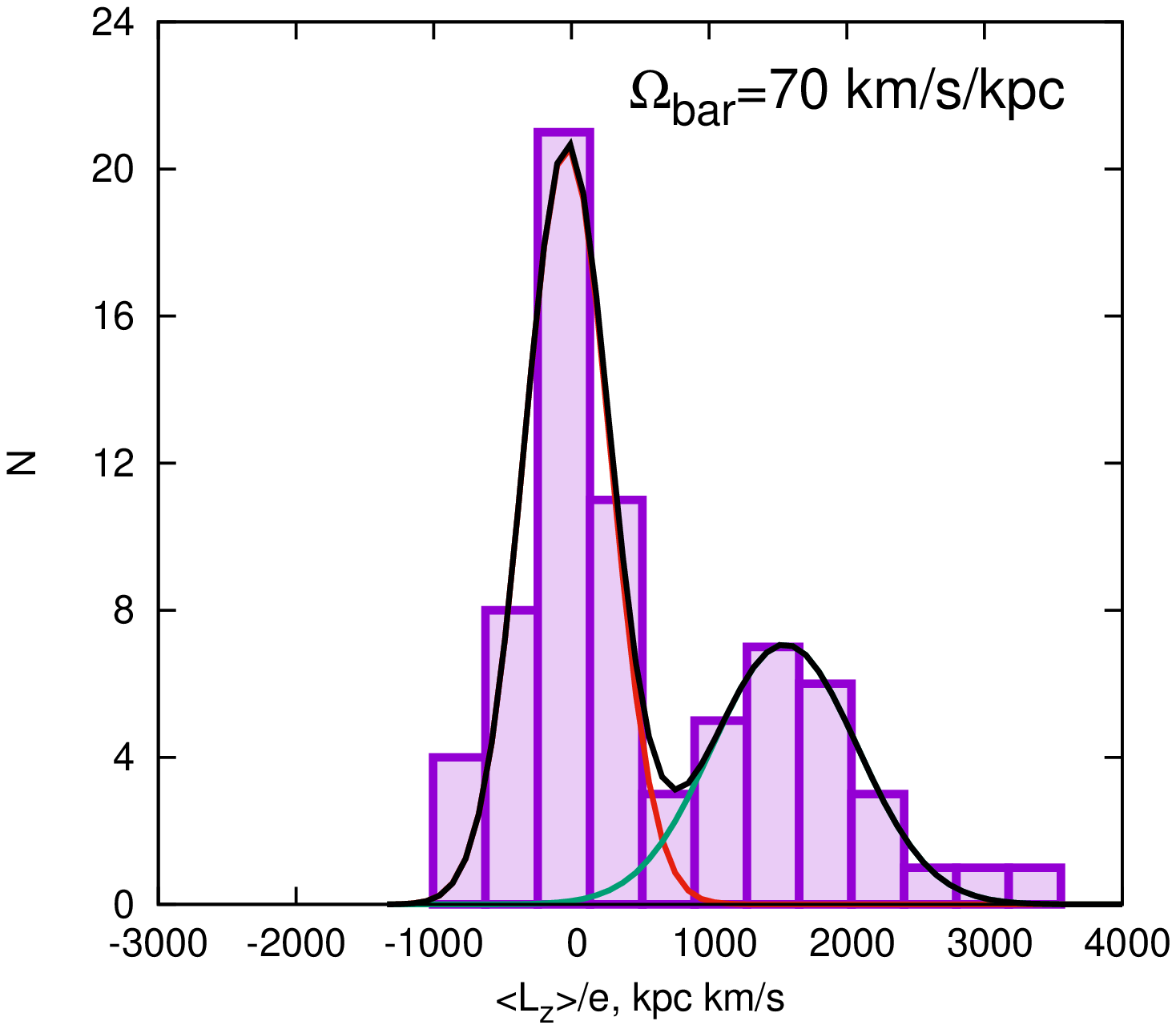}
\caption{ Distribution of the globular clusters by the ratio of the average angular momentum to eccentricity $<L_z>/e$ for different bar velocities: $\Omega_{bar}=31, 41, 55,$ and $70 $~km s$^{-1}$ kpc$^{-1}$.}
\label{fbar}
\end{center}}
\end{figure}

\noindent
Nonetheless, as expected, the inclusion of a rotating bar affected the separation of the GCs close to the Galactic center. As a result, the number of bulge objects decreased from 20,as  identified in the axisymmetric potential, to 16 for bar rotation velocities of 31, 41, and 55 km s$^{-1}$ kpc$^{-1}$ and  to 15 for  bar velocity of 70 km s$^{-1}$ kpc$^{-1}$.  As a consequence, NGC 6355, Pismis 26, Djorg 2, ESO 456-78 were dropped out of the list of the bar/bulge objects, and in the case of bar velocity of 70 km s$^{-1}$ kpc$^{-1}$,  also  NGC 6558. The selection of halo objects under the condition $Z_{max} > 6$ kpc, only  changed slightly. The halo list was depleted by just the cluster $E 3$, at a bar velocities of 41, 55, and 70 km s$^{-1}$ kpc$^{-1}$.
The separation of  the GCs into disk and halo objects inside the layer $|Z|<6 $ kpc has undergone the following changes depending on the bar velocity. At bar velocity of 31 km s$^{-1}$ kpc$^{-1}$, the number of disk objects became equal to 36 since  NGC 6541 was dropped from the previous list,  while ESO 456-78 and NGC 6717 were added. At bar velocity of 41 km s$^{-1}$ kpc$^{-1}$,  the number of disk objects became equal to 36 since  $E 3$ was dropped, and ESO 456-78 and NGC 6717 were added. At  bar velocity of 55 km s$^{-1}$ kpc$^{-1}$,  the number of disk objects became equal to 37, since  $E 3$ dropped out of the previous list, while ESO 456- 78, Djorg 2, and NGC 6717 were added.  Finally, at a bar velocity of 70 km s$^{-1}$ kpc$^{-1}$,  the number of disk objects became equal to 35 since $E 3$and  NGC 6541 were dropped from the previous list, while ESO 456-78 and NGC 6717  were added.
Thus, from 35 disk objects identified in the axisymmetric potential, 33 objects are stably preserved. The largest influence of the bar turned out to be on GCs $E 3$ and NGC 6541.
Overall, we found little dependence of the results of the separation of GCs on the bar rotation velocity.
The histograms of the GCs distributions by the $< L_z >/e$ parameter and the approximating Gaussian distributions for all the bar velocities are shown in Figure \ref{fbar}.

\section{Metallicities of the cluster subsystems}

It is known since long time that globular clusters with low metal abundances are distributed throughout the volume of the halo and have a low concentration toward the Galactic plane, while clusters with high metal abundances are concentrated toward the Galactic plane and are found predominantly at distances inside the solar radius. Zinn (1993) showed that clusters with metallicities $[Fe/H]\leq -0.8$ dex belong to the Milky Way halo, and have comparatively low rotational speeds and velocity dispersions while clusters with higher metal abundances belong to the disk subsystem.

Precise kinematical data of the Milky Way globular clusters allow to address the problem of the chemical abundances of the globular clusters in a new way: to solve the general problem of the inter-relation of the chemical abundances and the kinematical properties of the clusters. By assuming the separation of the clusters among the subsystems of the Galaxy, one can estimate the mean metallicities of the clusters that belong to the different subsystems of the Milky Way galaxy.

\begin{figure}[t]
{\begin{center}
   \includegraphics[width=0.3\textwidth,angle=0]{fig2.eps}
   \includegraphics[width=0.3\textwidth,angle=0]{fig3.eps}
   \includegraphics[width=0.3\textwidth,angle=0]{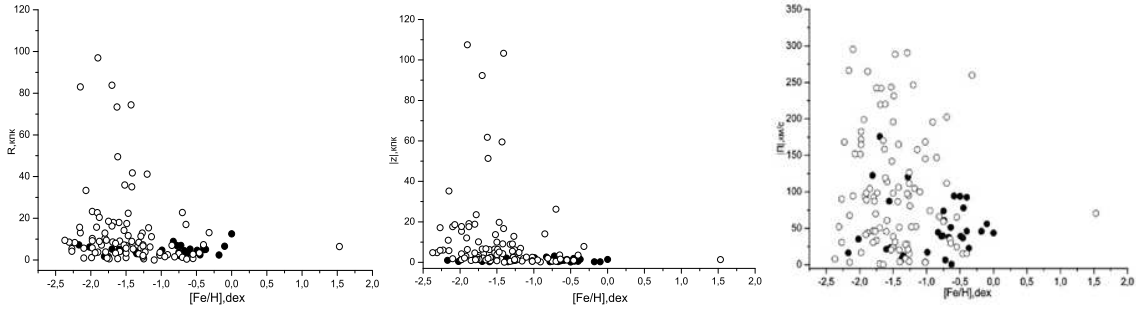}
\caption{ Distribution of the metallicities of globular clusters in the halo and thick disk subsystems as a function of distance from the Galactic center (left-hand panel); dependence of the metallicities of the globular clusters on distance from the Galactic plane Z (middle panel); dependence of the metallicities of the globular clusters on the radial velocities of halo  and disk (right-hand panel). In all pictures the halo and disk clusters
are marked by unfilled and filled symbols respectively.}
\label{f6}
\end{center}}
\end{figure}

In order to take the chemical compositions of globular clusters in different Galactic subsystems into account, we used cluster metallicities taken from Harris (2010), and for the Crater cluster from Laevens et al. (2014). Figure \ref{f6} on the left-hand panel shows the metallicity distribution for the halo and for the thick disk cluster subsystems as a function of distance from the Galactic center. The hollow and filled symbols correspond to the halo and disk clusters, respectively. Together with clusters with high metal abundances and disk kinematics, several low-metal clusters also display kinematics typical of the disk subsystem. This result can also be seen in Figure \ref{f6} in middle panel, which presents the metallicity distributions of halo and thick-disk globular clusters as a function of the distance from the plane of the disk, $Z$. This Figure shows that, together with clusters in the disk subsystem, globular clusters displaying halo kinematics (hollow symbols) are also concentrated toward the plane of the disk, probably reflecting some incompleteness of the sample.

The right-hand panel in Figure \ref{f6} shows metallicity distributions as the function of Galactocentric radial velocities of halo globular clusters (empty symbols) and disk globular clusters (filled symbols). This figure shows that a large fraction of disk globular clusters located
in the region with $[Fe/H] \geq -1.0$ dex, while most halo clusters have low metal abundances $[Fe/H] \leq -1.0$ dex. Virtually all clusters displaying halo kinematics have low metal abundances, and these clusters occupy a large range of radial velocities, right up to 300 km/s. We find that the mean metallicity for the disk globular clusters subsystem is equal to $[Fe/H] = -0.96\pm 0.11$ dex, while for the halo objects mean metallicity is equal to $[Fe/H] = -1.58\pm 0.05$  dex, and for the globular clusters in the region of the bar/bulge the mean metallicity is equal to $[Fe/H] = -0.95\pm 0.08$ dex. After the study of Forbes $\&$ Bridges (2010), retrograde motion has usually been considered to indicate that a globular cluster was accreted during the evolution of the Galaxy. If this view is correct, we should expect differences in the mean metallicities for accreted halo globular clusters exhibiting retrograde motion and clusters orbiting in the Galactic prograde direction, which are presumed to have formed  {\it in situ} during the evolution of the Galaxy. Cluster kinematics can be used to distinguish halo clusters in retrograde orbits and  clusters rotating in the prograde direction. To elucidate the likelihood of possible differences in the chemical compositions of these groups of globular clusters, we applied a statistical Student t-test. The mean metallicities of these two groups of clusters are statistically indistinguishable which does not support the suggestion that these two groups of clusters have different origins.

On the other hand, the application of  this test to the samples of globular clusters belonging to the disk and halo subsystems show statistically significant differences, indicating  that these two groups of clusters have different chemical evolution histories, and probably different origin. The mean metallicity of bar/bulge globular clusters statistically indistinguishable from mean metallicity of the disk globulars.

\section {Summary}
In this paper we presented a large suite of orbital  integrations of the Milky Way globular clusters with the aim of determining their membership to different Galactic components.
We summarise our main results as follows:

(i) We have used high quality proper motion data for nearly all known Milky Way globular clusters together with known distances and the line of sight velocities  to study their kinematical properties of the Milky Way components.  To assign clusters to the various subsystems of the Milky Way we used their orbital parameter (ratio $L_z/e$) distribution,  which was never applied before to the globular cluster system of the Milky Way. We have shown that our splitting algorithm is stable against different realisations of the Galactic potential, and both in the axisymmetric and non-axisymmetric cases.

(ii) The orbits of the clusters, and their orbital parameters clearly distinguish three groups of the clusters: those that belong to the bar/bulge of the Milky Way galaxy, the clusters, that belong to the thick Milky Way disk, and the clusters, belonging to the Milky Way halo.

(iii) The bar/bulge subsystem of the globular clusters has a relatively slow rotation of 49 $\pm$ 11 km s$^{-1}$, and close to zero radial velocity for the ensemble of the globular clusters within the bar/bulge region

(iv) The clusters distinguished by their kinematical properties as the members of the thick disk, have rotational velocity of 179 $\pm$ 6 km s$^{-1}$ consistent with the recent estimate of the rotational velocity of metal-weak stellar thick disk population 182 $\pm$ 6 km s$^{-1}$ (Tian et al. 2019).

The estimated lag in the rotational velocity of the globular clusters thick disk and the rotational velocity of LSR is 65 $\pm$ 6 km s$^{-1}$ roughly consistent with the estimate of the lag of the rotational velocity between the stellar thick and the thin disks found by Pasetto et al. (2012).

Our estimate of the rotational velocity of the thick disk globular clusters differs from the recent estimate of the disk globular clusters rotational velocity obtained by Posti $\&$ Helmi (2018) equal to 220 km s$^{-1}$ but is consistent with recent estimate of the rotational velocity of the metal-weak thick disk stellar population (Tian et al. 2019).

(v) We confirm previous findings that the subsystem of the globular clusters belonging to the Milky Way halo practically has rotation close to zero 1 $\pm$ 4 km s$^{-1}$. We find contrary to the conclusion of Posti $\&$ Helmi (2018) , that the halo globular clusters have a prograde rotation, and its rotational velocity is consistent with the estimated rotational velocity of local stellar halo of 27$^{+4}_{-5}$ km s$^{-1}$ (Tian et al. 2019).

(vi) We have determined the mean metallicities for globular clusters with well defined kinematic parameters and orbits in different subsystems of the Milky Way. The mean metallicity of globular clusters of the disk subsystem was found to be  $[Fe/H] = -0.96\pm 0.11$ dex, and of the halo subsystem to be $[Fe/H] = -1.58\pm 0.05$ dex. The mean metallicity of globular clusters associated with the bar/bulge was found to be $[Fe/H] = -0.95\pm 0.08$ dex.

\section*{Acknowledgments}
VK acknowledges financial support by Southern Federal University, 2020 (Ministry of Science and Higher Education of the Russian Federation).
We acknowledge the anonymous referee for the valuable comments and suggestions.

 \medskip\subsubsection*{REFERENCES}

{\small
\quad   Allen, C., $\&$ Santillan, A. 1991, RMxAA, 22, 255

 Antoja, T., Helmi, A., Dehen, W., et al. 2014, A$\&$A, 563, id.60, 17 pp.

 Bajkova, A. T.,$\&$ Bobylev, V. V.  2016, AstL, 42, 567

 Bajkova, A. T.,$\&$ Bobylev, V. V.  2017, OAst,  26, 72

 Baumgardt, H., Hilker, M., Sollima, A., $\&$ Bellini, A. 2019, MNRAS, 482, 5138

 Bhattacharjee, P., Chaudhury, S., $\&$ Kundu, S. 2014, ApJ, 785, 63

 Binney, J., $\&$ Tremaine, S. 1987, Galactic Dynamics. Princeton Univ. Press, Princeton, NJ

 Binney, J., $\&$  Wong, L. K. 2017, MNRAS, 467, 2446

 Bobylev, V. V., $\&$  Bajkova, A. T. 2016, AstL, 42, 228

 Debattista, V. P., Gerhard, O., Sevenster, M. N. 2002, MNRAS, 334, 355

 Forbes, D. A.,  $\&$ Bridges, T. 2010, MNRAS, 404, 1203

 Forbes, D. A. 2010, arXiv: 2002.01512

 Gaia Collaboration, Helmi, A., et al. 2018, A$\&$A, 616,12

 Harris, W. 2010, arXiv: 1012.3224

 Hernquist, L. 1990, ApJ, 356, 359

 Horta, D., Schiavon, R. P., Mackereth, J. T., et al. 2020, arXiv: 2001.03177

 Howard, C. D., Rich, R. M., Reitzel, D. B., et al. 2008, ApJ, 688, 1060

 Irrgang, A., Wilcox, B., Tucker, E., $\&$ Schiefelbein, L. 2013, A$\&$A, 549, 137

 Kunder, A., Koch, A., Rich, R. M., et al. 2012, AJ, 143, id 57, 14 pp.

 Laevens, B. P. M.,  Martin, N. F., Sesar, B., et al. 2014, ApJ, 786, L3

 Mackey, A. D., $\&$ Gilmore, G. F. 2004, MNRAS, 355, 504

 Massari, D., Koppelman H. H., $\&$  Helmi, A. 2019, A$\&$A, 630, L4

 Miyamoto, M., $\&$ Nagai, R.  1975, PASJ, 27, 533

 Myeong, G. C., Vasiliev, E., Iorio, G., Evans, N. W., $\&$ Belokurov, V. 2019, MNRAS, 488, 1235

 Navarro, J. F., Frenk, C.S., $\&$ White, S.D.M. 1997, ApJ, 490, 493

 Ness, M., Freeman, K., Athanassoula, E., et al. 2012, ApJ, 756, id 22, 8 pp.

 Palou\v{s}, J., Jungwiert, B.,  $\&$ Kopeck\'y, J. 1993, A$\&$A, 274, 189

 Pasetto, E. K., Grebel, T., Zwitter, C., et al. 2012, ApJ, 547A, 71

 P\'erez-Villegas, A., Barbuy, B., Kerber, L., et al. 2020, MNRAS, 491, 3251

 Piatti, A. E. 2019, ApJ, 882, 98

 Plummer, H. C. 1911, MNRAS, 71, 460

 Posti, L., Helmi, A., Veljanoski, J., $\&$ Breddels, M. A. 2018, A$\&$A, 615, A70

 Posti, L., $\&$ Helmi, A. 2019, A$\&$A, 621, 56

 Sanders, J. L., Smith, L., Evans, N. W., Lucas, P. 2019, MNRAS, 487, 5188

 Sch\"onrich, R., Binney, J., $\&$ Dehnen, W. 2010, MNRAS,  403, 1829

 Sohn, S. T., Watkins, L. L., Fardal, M. A., et al. 2018, ApJ, 862, 52

 Tian, H., Liu, C., Xu, Y., $\&$ Xue, X. 2019, ApJ, 871, 184

 Vasiliev, E. 2019, MNRAS, 484, 2832

 Wegg, C., $\&$ Gerhard, O. 2013, MNRAS, 435, 1874

 Wilkinson, M. I., $\&$ Evans, N. W. 1999, MNRAS, 310, 645

 Zoccali, M., Gonzales, O. A., Vasquez, S., et al. 2014, A$\&$A, 562, id A66, 11 pp.

 Zinn, R. 1993, ASPC, 48, 38

}

\section*{Appendix}

\centerline{{\bf Table A1.} Orbit parameters of globular clusters obtained in the NFWBB axisymmetric potential}
		\label{t:A1}
		\begin{small} \begin{center}\begin{tabular}{|l|c|c|c|c|c|c|c|c|c|c|c|c|r|}\hline
Name & $X$ & $Y$ & $Z$ & $\Pi$ & $\Theta$ & $W$ & $e$ & $Z_{max}$ & $R_{max}$& $r_{max}$& $L_z$&E& GS \\
   &  kpc &  kpc & kpc &km/s & km/s& km/s&  &kpc & kpc& kpc & kpc km/s& (km/s)$^2$&  \\\hline
NGC 104  &   6.4&  -2.6&  -3.2&   6& 192&  45&0.16& 3.4& 7.5& 7.7&  1328&   -126288&D\\
NGC 288  &   8.4&   0.1&  -8.9&   9& -42&  51&0.80&10.3&12.3&12.3&  -349&   -116280&H\\
NGC 362  &   5.2&  -5.1&  -6.2& 127&  -1& -70&0.98&10.4&12.3&12.3&   -10&   -121290&H\\
Whiting 1  &  22.3&   4.7& -26.2&-208& 109&   8&0.53&60.9&66.8&67.0&  2494&    -42265&H\\
NGC 1261 &   8.2& -10.0& -12.8& -95& -19&  69&0.94&16.8&21.1&21.1&  -249&    -91385&H\\
Pal 1   &  15.1&   8.0&   3.6&  42& 214& -20&0.13& 4.5&19.1&19.1&  3646&    -78086&D\\
E 1     &  24.8& -80.1& -92.3&  11&-144&  96&0.31&135.2&231.9&237.9&-12100&    -15914&H\\
Eridanus  &  61.5& -41.8& -59.5& -90& -28& 133&0.87&140.9&145.9&174.5& -2064&    -23988&H\\
Pal 2   &  34.8&   4.4&  -4.3&-108&  11&   4&0.97&13.7&40.9&41.0&   373&    -63896&H\\
NGC 1851 &  12.6&  -8.9&  -6.9& 105&  -1& -82&0.99&17.0&20.1&20.1&   -22&    -94063&H\\
NGC 1904 &  15.9&  -8.3&  -6.3&  46&  12&   4&0.97& 8.8&19.7&19.7&   211&    -96192&H\\
NGC 2298 &  12.6&  -9.5&  -3.0& -92& -32&  77&0.87&11.4&18.0&18.0&  -500&    -98716&H\\
NGC 2419 &  83.0&  -0.5&  35.2&  -5&  47& -58&0.68&58.1&83.1&91.8&  3907&    -35896&H\\
Pyxis   &  14.2& -38.7&   4.8&-247& -29& 187&0.77&283.5&175.6&324.2& -1205&    -15098&H\\
NGC 2808 &   6.3&  -9.2&  -1.9&-157&  41&  28&0.87& 4.9&14.2&14.4&   457&   -111797&H\\
E 3     &   5.4&  -7.1&  -2.6&  44& 251& 104&0.18& 5.8&13.0&13.1&  2240&    -97684&D\\
Pal 3   &  42.6& -59.7&  61.8&-147&  89&  67&0.39&134.7&153.7&173.4&  6495&    -20091&H\\
NGC 3201 &   7.7&  -4.8&   0.8&-114&-301& 151&0.52&10.8&26.2&26.3& -2728&    -75483&H\\
Pal 4   &  39.7& -12.9& 103.3& -25& -33&  56&0.75&110.9&114.1&116.5& -1394&    -31065&H\\
Crater  &   0.2& -97.0& 107.5&-101& -63&  63&0.14&125.4&156.1&156.1& -6104&    -18859&H\\
NGC 4147 &   9.6&  -4.1&  18.8&  47&  -3& 128&0.97&26.0&25.8&26.4&   -27&    -81117&H\\
NGC 4372 &   5.4&  -4.9&  -1.0&  16& 133&  67&0.42& 2.1& 7.3& 7.3&   962&   -139570&D\\
Rup 106  &  -2.4& -17.8&   4.3&-242&  91&  32&0.78&23.7&37.9&37.9&  1640&    -64986&H\\
NGC 4590 &   4.2&  -7.2&   6.1&-169& 293&  16&0.54&17.9&29.0&29.9&  2453&    -70495&H\\
NGC 4833 &   4.7&  -5.4&  -0.9& 105&  40& -42&0.83& 3.5& 8.0& 8.0&   286&   -144428&H\\
NGC 5024 &   5.5&  -1.5&  17.6& -95& 141& -72&0.43&21.4&22.1&22.3&   797&    -80241&H\\
NGC 5053 &   5.3&  -1.4&  17.1& -89& 134&  35&0.27&17.4&17.8&18.0&   727&    -85221&H\\
NGC 5139 &   5.1&  -3.9&   1.4& -70& -72& -80&0.73& 2.8& 7.4& 7.4&  -462&   -147850&H\\
NGC 5272 &   6.8&   1.3&  10.0& -38& 143&-134&0.51&13.3&15.7&15.9&   994&    -98226&H\\
NGC 5286 &   0.7&  -8.6&   2.2&-220& -44&  10&0.89& 7.0&13.7&13.7&  -375&   -113332&H\\
NGC 5466 &   5.0&   3.0&  15.4& 172&-141& 226&0.80&49.4&53.6&53.7&  -820&    -52693&H\\
NGC 5634 &  -7.4&  -5.0&  19.1& -45&  39& -26&0.82&19.7&21.1&21.6&   346&    -89143&H\\
NGC 5694 & -18.1& -14.6&  17.7&-185& -44&-166&0.93&46.2&69.9&70.4& -1019&    -45290&H\\
IC 4499 &  -2.4& -14.0&  -6.6&-245& -75& -62&0.65&27.1&29.8&30.0& -1066&    -72235&H\\
NGC 5824 & -18.1& -13.7&  12.1& -41& 105&-183&0.45&31.6&37.3&37.4&  2393&    -59779&H\\
Pal 5   &  -7.9&   0.2&  16.7& -54& 160& -10&0.27&17.2&18.8&18.9&  1260&    -83194&H\\
NGC 5897 &  -2.0&  -3.2&   6.3&  88&  97&  90&0.64& 7.6& 8.6& 8.7&   362&   -131771&H\\
NGC 5904 &   3.2&   0.4&   5.5&-290& 126&-181&0.82&21.1&23.0&23.3&   402&    -85576&H\\
NGC 5927 &   1.9&  -4.2&   0.7& -39& 233&   5&0.11& 0.7& 5.2& 5.2&  1077&   -148662&D\\
NGC 5946 &  -0.6&  -5.7&   0.8&  35&  25& 108&0.89& 4.3& 5.8& 5.9&   141&   -158006&H\\
ESO 224-8 &  -7.8&  -9.9&   1.5& -43& 259&   0&0.18& 1.9&17.0&17.0&  3245&    -85205&D\\
NGC 5986 &  -1.0&  -4.0&   2.4&  62&  23& -14&0.92& 3.8& 4.9& 5.0&    94&   -168505&H\\
\hline
				\end{tabular}\end{center}
			\end{small}
	%\end{table}

\newpage

\centerline{{\bf Table A1.}-- Continued. Orbit parameters of globular clusters obtained in the NFWBB axisymmetric potential}
		\label{t:A1}
		\begin{small} \begin{center}\begin{tabular}{|l|c|c|c|c|c|c|c|c|c|c|c|c|r|}\hline
Name & $X$ & $Y$ & $Z$ & $\Pi$ & $\Theta$ & $W$ & $e$ & $Z_{max}$ & $R_{max}$& $r_{max}$& $L_z$&E& GS \\
   &  kpc &  kpc & kpc &km/s & km/s& km/s&  &kpc & kpc& kpc & kpc km/s& (km/s)$^2$&  \\\hline
FSR 1716 &   0.1&  -4.8&  -0.3&  87& 228&-139&0.28& 2.5& 6.9& 7.0&  1089&   -137191&D\\
Pal 14  & -41.4&  27.3&  51.4& 117&  16& 131&0.97&66.4&127.5&133.9&   794&   -29458&H\\
BH 184  &   1.5&  -4.1&  -0.4&  40& 121& -89&0.47& 1.5& 4.6& 4.7&   531&   -168579&D\\
NGC 6093 &  -1.1&  -1.2&   3.4&  33&  16& -61&0.97& 3.7& 4.6& 4.6&    25&   -176933&H\\
NGC 6121 &   6.2&  -0.3&   0.6& -52&  10&  -9&0.97& 4.2& 6.5& 6.5&    60&   -159021&H\\
NGC 6101 &  -2.7& -10.0&  -4.2& -12&-314&-195&0.61&24.3&42.2&44.2& -3236&    -56722&H\\
NGC 6144 &  -0.2&  -1.2&   2.4& -69&-196&  43&0.21& 2.9& 2.9& 3.3&  -239&   -172662&H\\
NGC 6139 &  -1.3&  -3.0&   1.2&  -1&  76& 137&0.54& 2.7& 3.4& 3.6&   248&   -176480&H\\
Terzan 3 &   0.5&  -2.1&   1.3& -61& 206&  99&0.18& 1.9& 3.0& 3.2&   440&   -175324&D\\
NGC 6171 &   2.4&   0.4&   2.5&  -4&  78& -65&0.72& 2.8& 3.7& 3.8&   191&   -178959&H\\
ESO 452-11 &   0.3&  -1.1&   1.8& -24& -13&-104&0.98& 2.2& 2.9& 2.9&   -16&   -202684&H\\
NGC 6205 &   5.5&   4.6&   4.7&  20& -26& -80&0.79& 7.8& 8.6& 8.6&  -187&   -134416&H\\
NGC 6229 &   1.8&  22.3&  19.8&  30&   6&  51&0.97&27.8&30.9&31.0&   135&    -74434&H\\
NGC 6218 &   4.2&   1.2&   2.1&  -9& 135& -82&0.37& 2.7& 4.9& 5.0&   581&   -158135&D\\
FSR 1735 &  -1.8&  -3.8&  -0.3&-102&  -5& 201&0.69& 5.3& 5.2& 5.3&   -22&   -157538&H\\
NGC 6235 &  -2.9&  -0.2&   2.7& 159& 197& -40&0.39& 4.5& 5.9& 6.2&   570&   -145439&D\\
NGC 6254 &   4.4&   1.1&   1.7& -87& 134&  48&0.42& 2.4& 5.2& 5.2&   606&   -157472&D\\
NGC 6256 &  -1.8&  -2.2&   0.6&-170&  28&  97&0.92& 3.1& 4.2& 4.2&    79&   -181873&H\\
Pal 15  & -30.6&  13.3&  18.6& 154&  -5&  50&0.96&50.8&49.4&54.6&  -165&    -53233&H\\
NGC 6266 &   1.6&  -0.8&   0.9&  42& 122&  69&0.62& 1.0& 2.5& 2.5&   215&   -205537&B\\
NGC 6273 &  -0.4&  -0.5&   1.5& -98&-240& 179&0.59& 3.5& 3.8& 3.9&  -144&   -172941&H\\
NGC 6284 &  -6.8&  -0.4&   2.7&  14&  -2& 112&0.83& 7.3& 7.0& 7.5&   -16&   -142332&H\\
NGC 6287 &  -0.9&   0.0&   1.8&-301& -64&  79&0.75& 5.2& 5.3& 5.3&   -60&   -159009&H\\
NGC 6293 &  -1.1&  -0.4&   1.3&-152& -80&-158&0.91& 2.5& 3.6& 3.6&   -93&   -191358&H\\
NGC 6304 &   2.4&  -0.4&   0.6&  79& 191&  72&0.29& 1.0& 3.2& 3.3&   474&   -183132&D\\
NGC 6316 &  -2.0&  -0.5&   1.1& 103&  51&  86&0.77& 1.5& 2.9& 3.0&   106&   -197316&B\\
NGC 6341 &   5.8&   6.3&   4.8&  53&  13&  95&0.94& 9.4&10.7&10.7&   108&   -125312&H\\
NGC 6325 &   0.6&   0.1&   1.1& -81&-181&  79&0.12& 1.2& 1.2& 1.4&  -107&   -212097&H\\
NGC 6333 &   0.6&   0.8&   1.5& -89& 346&  68&0.74& 4.1& 6.4& 6.4&   327&   -151409&H\\
NGC 6342 &  -0.1&   0.7&   1.5& -25& 164& -32&0.31& 1.5& 1.6& 1.7&   117&   -207010&H\\
NGC 6356 &  -6.5&   1.7&   2.7&  47& 107& 110&0.52& 4.5& 7.8& 7.9&   713&   -136303&D\\
NGC 6355 &  -0.9&  -0.1&   0.9&-207&-110& 144&0.56& 1.9& 2.0& 2.2&   -95&   -199192&B\\
NGC 6352 &   3.0&  -1.8&  -0.7&  42& 226&   7&0.13& 0.7& 4.1& 4.1&   794&   -163864&D\\
IC 1257 & -14.8&   6.9&   6.5& -45& -50& -20&0.82& 6.7&18.1&18.1&  -817&    -98556&H\\
Terzan 2 &   0.8&  -0.5&   0.3&-120& -47& -44&0.86& 0.7& 1.2& 1.2&   -44&   -242360&B\\
NGC 6366 &   5.1&   1.1&   1.0&  94& 134& -62&0.45& 2.0& 5.8& 5.8&   699&   -153378&D\\
Terzan 4 &   1.1&  -0.5&   0.2&  15&  75&  97&0.68& 0.8& 1.3& 1.3&    92&   -234494&B\\
BH 229  &   0.1&  -0.4&   0.3&   7& -55&-287&0.49& 0.8& 0.8& 0.8&   -21&   -249808&B\\
NGC 6362 &   2.3&  -4.1&  -2.3&  17& 124& 100&0.37& 3.3& 5.2& 5.3&   583&   -153468&D\\
Liller 1 &   0.1&  -0.7&  -0.0& 107& -56&  25&0.81& 0.2& 0.8& 0.8&   -42&   -261395&B\\
NGC 6380 &  -2.4&  -1.9&  -0.6& -62& -35&  14&0.89& 2.1& 3.4& 3.4&  -105&   -194646&H\\
Terzan 1 &   1.6&  -0.3&   0.1& -73&  63&   5&0.79& 1.0& 1.8& 1.8&   102&   -224803&B\\
Pismis 26 &   0.2&  -1.3&  -0.5&-112& 204& 200&0.56& 1.8& 3.2& 3.2&   271&   -186876&B\\
NGC 6388 &  -1.2&  -2.5&  -1.1& -66& -94& -16&0.69& 1.6& 3.5& 3.5&  -257&   -190150&H\\
NGC 6402 &  -0.1&   3.3&   2.4& -20&  48&  23&0.88& 2.8& 4.8& 4.8&   158&   -176457&H\\
NGC 6401 &  -2.3&   0.6&   0.8& -30&-254& 161&0.31& 2.3& 4.4& 4.5&  -595&   -161761&H\\
NGC 6397 &   6.2&  -0.8&  -0.5&  35& 127&-121&0.40& 2.9& 6.5& 6.5&   796&   -144538&D\\
\hline
				\end{tabular}\end{center}
			\end{small}
%	\end{table}}

\newpage

\centerline{{\bf Table A1.}-- Continued. Orbit parameters of globular clusters obtained in the NFWBB axisymmetric potential}
		\label{t:A1}
		\begin{small} \begin{center}\begin{tabular}{|l|c|c|c|c|c|c|c|c|c|c|c|c|r|}\hline
Name & $X$ & $Y$ & $Z$ & $\Pi$ & $\Theta$ & $W$ & $e$ & $Z_{max}$ & $R_{max}$& $r_{max}$& $L_z$&E& GS \\
   &  kpc &  kpc & kpc &km/s & km/s& km/s&  &kpc & kpc& kpc & kpc km/s& (km/s)$^2$&  \\\hline
Pal 6   &   2.5&   0.2&   0.2&-191&  21& 153&0.95& 3.3& 4.5& 4.5&    52&   -179351&H\\
NGC 6426 &  -9.2&   9.3&   5.8&-112&  93& -26&0.67& 7.0&16.5&16.6&  1216&   -100666&H\\
Djorg 1  &  -1.0&  -0.5&  -0.4&-252& 315&  28&0.76& 2.6& 5.9& 6.0&   351&   -161437&H\\
Terzan 5 &   1.4&   0.5&   0.2&  84&  70& -31&0.77& 0.8& 1.7& 1.7&   104&   -226313&B\\
NGC 6440 &  -0.1&   1.1&   0.6&  91& -42& -39&0.78& 1.1& 1.4& 1.4&   -49&   -231796&B\\
NGC 6441 &  -3.2&  -1.3&  -1.0&  16&  66& -21&0.66& 1.4& 3.5& 3.6&   228&   -186312&H\\
Terzan 6 &   1.5&  -0.2&  -0.2&-138& -51&  42&0.86& 1.3& 2.0& 2.0&   -77&   -220185&B\\
NGC 6453 &  -3.2&  -0.9&  -0.8&-105&  38&-159&0.61& 3.6& 3.8& 3.9&   129&   -172906&H\\
NGC 6496 &  -2.6&  -2.3&  -2.0& -37& 320& -60&0.42& 4.0& 9.0& 9.1&  1111&   -126509&D\\
Terzan 9 &   1.2&   0.5&  -0.2& -50&  22& -53&0.92& 1.0& 1.4& 1.4&    29&   -236057&B\\
Djorg 2  &   2.0&   0.3&  -0.3& 161& 155& -45&0.57& 0.4& 3.2& 3.2&   316&   -192751&B\\
NGC 6517 &  -1.6&   3.5&   1.3&  55&  33& -34&0.90& 2.9& 4.5& 4.5&   127&   -179281&H\\
Terzan 10 &  -2.1&   0.8&  -0.3& 231&  87& 239&0.79& 4.7& 5.9& 5.9&   193&   -155422&H\\
NGC 6522 &   0.6&   0.1&  -0.5&  34&  92&-189&0.67& 0.9& 1.2& 1.2&    58&   -238519&B\\
NGC 6535 &   2.4&   3.0&   1.3&  93& -83&  45&0.64& 1.6& 4.5& 4.6&  -320&   -173024&H\\
NGC 6528 &   0.4&   0.2&  -0.6&-197& 113& -32&0.60& 0.9& 1.0& 1.0&    51&   -241883&B\\
NGC 6539 &   1.1&   2.8&   0.9&   1& 118& 172&0.30& 2.5& 3.2& 3.4&   347&   -174387&D\\
NGC 6540 &   3.0&   0.3&  -0.3&  13& 148&  57&0.32& 0.5& 3.0& 3.1&   448&   -187517&D\\
NGC 6544 &   5.3&   0.3&  -0.1&   6&   6& -90&0.98& 4.3& 5.7& 5.7&    31&   -166630&H\\
NGC 6541 &   1.1&  -1.4&  -1.4& 123& 192&-112&0.50& 2.4& 3.8& 3.8&   334&   -174968&D\\
ESO 280-06 & -12.0&  -4.7&  -4.6&  31&  16& -84&0.88&12.6&14.2&14.2&   210&   -109791&H\\
NGC 6553 &   2.3&   0.6&  -0.3&  45& 245&  -5&0.19& 0.3& 3.3& 3.3&   588&   -179839&D\\
NGC 6558 &   0.9&   0.0&  -0.8& 187&  93&  15&0.72& 1.4& 1.7& 1.7&    87&   -217130&B\\
Pal 7   &   3.3&   2.0&   0.6& -74& 270&  27&0.26& 0.8& 6.0& 6.0&  1042&   -147032&D\\
Terzan 12 &   3.5&   0.7&  -0.2& -94& 172&  97&0.33& 1.3& 4.3& 4.4&   625&   -167749&D\\
NGC6569 &  -2.5&   0.1&  -1.3& -40& 174&  25&0.23& 1.3& 2.9& 3.0&   440&   -182361&D\\
ESO 456-78 &   1.8&   0.4&  -0.6&  71& 199&-139&0.34& 1.3& 2.9& 2.9&   373&   -186632&B\\
NGC 6584 &  -4.0&  -4.0&  -3.8& 197&  98&-238&0.83&12.9&17.9&18.0&   556&    -98089&H\\
NGC 6624 &   0.5&   0.4&  -1.1& -29&  60&-119&0.78& 1.3& 1.5& 1.5&    37&   -226410&B\\
NGC 6626 &   2.9&   0.7&  -0.5& -28&  57& -93&0.75& 1.9& 3.1& 3.1&   169&   -193067&H\\
NGC 6638 &  -0.9&   1.3&  -1.1&  68&  14&  26&0.96& 1.8& 2.4& 2.4&    22&   -212068&H\\
NGC 6637 &  -0.4&   0.3&  -1.6&  35&  91&  85&0.94& 1.7& 2.4& 2.4&    40&   -212500&H\\
NGC 6642 &   0.4&   1.4&  -0.9& 112&  25& -50&0.94& 1.6& 2.2& 2.2&    36&   -215457&B\\
NGC 6652 &  -1.5&   0.3&  -2.0& -54&  28& 176&0.96& 3.1& 4.2& 4.2&    42&   -183595&H\\
NGC 6656 &   5.2&   0.6&  -0.4& 176& 201&-143&0.53& 3.5& 9.8& 9.8&  1044&   -125471&D\\
Pal 8   &  -4.0&   3.1&  -1.5& -21& 117& -29&0.51& 1.7& 5.5& 5.6&   593&   -160020&D\\
NGC 6681 &  -0.5&   0.4&  -1.9& 221&  55&-176&0.74& 4.4& 4.5& 4.5&    36&   -167768&H\\
NGC 6712 &   2.1&   3.0&  -0.5& 146&  26&-146&0.94& 3.9& 5.5& 5.5&    93&   -168544&H\\
NGC 6715 & -17.3&   2.5&  -6.4& 232&  49& 210&0.58&51.4&52.3&56.4&   850&    -48159&H\\
NGC 6717 &   1.5&   1.6&  -1.3& -10& 116&  25&0.59& 1.4& 2.8& 2.8&   251&   -195762&H\\
NGC 6723 &  -0.0&   0.0&  -2.6& 100&-178& -38&0.26& 3.1& 2.7& 3.1&    -2&   -175356&H\\
NGC 6749 &   1.9&   4.7&  -0.3& -23& 110&   2&0.53& 0.3& 5.1& 5.1&   556&   -167068&D\\
NGC 6752 &   5.0&  -1.4&  -1.7& -23& 179&  59&0.23& 2.0& 5.6& 5.7&   931&   -147164&D\\
NGC 6760 &   2.3&   4.3&  -0.5&  92& 147& -15&0.44& 0.5& 5.5& 5.6&   724&   -158732&D\\
NGC 6779 &   4.0&   8.3&   1.4& 155& -15&  99&0.96& 9.4&12.4&12.4&  -135&   -119696&H\\
Terzan 7 & -13.1&   1.3&  -7.8& 260&  25& 185&0.54&41.3&41.0&42.9&   335&    -56545&H\\
Pal 10  &   4.7&   4.7&   0.3& -56& 186&  21&0.27& 0.4& 7.0& 7.0&  1234&   -138495&D\\
\hline
				\end{tabular}\end{center}
			\end{small}
%	\end{table}}
\newpage

\centerline{{\bf Table A1.} -- Continued. Orbit parameters of globular clusters obtained in the NFWBB axisymmetric potential}
		\label{t:A1}
		\begin{small} \begin{center}\begin{tabular}{|l|c|c|c|c|c|c|c|c|c|c|c|c|r|}\hline
Name & $X$ & $Y$ & $Z$ & $\Pi$ & $\Theta$ & $W$ & $e$ & $Z_{max}$ & $R_{max}$& $r_{max}$& $L_z$&E& GS \\
   &  kpc &  kpc & kpc &km/s & km/s& km/s&  &kpc & kpc& kpc & kpc km/s& (km/s)$^2$&  \\\hline
Arp 2   & -18.1&   4.0& -10.1& 243&  68& 181&0.57&55.6&58.7&65.1&  1256&    -43628&H\\
NGC 6809 &   3.4&   0.8&  -2.1&-199&  76& -55&0.66& 4.7& 5.7& 5.7&   266&   -154417&H\\
Terzan 8 & -15.5&   2.4& -10.9& 269&  37& 161&0.57&49.9&50.9&58.5&   584&    -46785&H\\
Pal 11  &  -2.7&   6.8&  -3.6& -16& 139&  -7&0.40& 3.6& 8.1& 8.2&  1013&   -132057&D\\
NGC 6838 &   6.1&   3.3&  -0.3&  39& 204&  39&0.18& 0.7& 7.3& 7.3&  1423&   -132307&D\\
NGC 6864 &  -9.4&   6.5&  -9.1& -99&  18&  48&0.94&12.8&16.4&16.4&   209&   -103615&H\\
NGC 6934 &  -0.8&  11.7&  -5.0&-289& 103& 122&0.88&15.4&40.8&40.9&  1204&    -63341&H\\
NGC 6981 &  -3.4&   8.2&  -9.2&-154&  -4& 170&0.97&15.9&22.3&22.3&   -35&    -89723&H\\
NGC 7006 &  -8.9&  34.9& -13.7&-140& -33&  86&0.90&35.5&55.8&56.3& -1170&    -52073&H\\
NGC 7078 &   4.4&   8.4&  -4.8&   8& 118& -27&0.50& 4.9&10.5&10.6&  1119&   -119947&D\\
NGC 7089 &   2.7&   7.5&  -6.7& 170& -18&-173&0.94&13.0&18.8&18.8&  -147&    -97640&H\\
NGC 7099 &   3.4&   2.5&  -5.9& -32& -55& 109&0.78& 7.0& 8.1& 8.2&  -234&   -137480&H\\
Pal 12  &  -2.7&   6.5& -14.0& 146& 304& 115&0.65&61.9&68.2&72.0&  2142&    -41826&H\\
Pal 13  &   7.3&  19.1& -17.6& 268& -78& -78&0.83&78.3&84.3&87.1& -1586&    -38661&H\\
NGC 7492 &   1.3&   9.4& -23.5& -87& -13&  64&0.80&27.8&28.0&28.1&  -120&    -77041&H\\\hline
				\end{tabular}\end{center}
			\end{small}
%	\end{table}}
\newpage

\rotatebox{90}{
	\label{t:A2}
		\begin{minipage}{1.5\linewidth}
\begin{small}
 \begin{center}
{\bf Table A2.} Orbit parameters of globular clusters in barred NFWBB potential
\medskip
 \begin{tabular}{|l|ccc|ccc|ccc|ccc|}\hline
Name&\multicolumn{3}{|c|}{$\Omega_{bar}=31$ km s$^{-1}$ kpc$^{-1}$}&\multicolumn{3}{|c|}{$\Omega_{bar}=41$ km s$^{-1}$ kpc$^{-1}$}&
\multicolumn{3}{|c|}{$\Omega_{bar}=55$ km s$^{-1}$ kpc$^{-1}$}&\multicolumn{3}{|c|}{$\Omega_{bar}=70$ km s$^{-1}$ kpc$^{-1}$}\\
 &$e$&$[L_z^{max},L_z^{min}]$&$<L_z>$&$e$&$[L_z^{max},L_z^{min}]$&$<L_z>$&
 $e$&$[L_z^{max},L_z^{min}]$&$<L_z>$&$e$&$[L_z^{max},L_z^{min}]$&$<L_z>$\\
 &  &kpc km s$^{-1}$&kpc km s$^{-1}$&  &kpc km s$^{-1}$&kpc km s$^{-1}$&  &kpc km s$^{-1}$&kpc km s$^{-1}$&  &kpc km s$^{-1}$&kpc km s$^{-1}$\\\hline
NGC 104  &0.17&[  1498,  1315]&  1397&0.14&[  1334,  1298]&  1319&0.14&[  1342,  1309]&  1328&0.16&[  1340,  1322]&  1332\\
Pal 1   &0.14&[  3651,  3637]&  3645&0.13&[  3648,  3644]&  3647&0.13&[  3648,  3645]&  3647&0.13&[  3647,  3645]&  3646\\
NGC 2808 &0.86&[   471,   445]&   458&0.86&[   461,   448]&   456&0.87&[   461,   449]&   455&0.86&[   471,   451]&   461\\
E 3     &0.24&[  2259,  2231]&  2243&0.24&[  2249,  2229]&  2240&0.23&[  2245,  2235]&  2240&0.21&[  2242,  2235]&  2239\\
NGC 4372 &0.39&[   992,   957]&   972&0.42&[   963,   835]&   901&0.38&[   964,   936]&   951&0.41&[  1010,   903]&   946\\
NGC 4833 &0.83&[   293,   275]&   285&0.83&[   292,   273]&   282&0.83&[   290,   263]&   277&0.83&[   304,   277]&   291\\
NGC 5139 &0.68&[  -476,  -448]&  -461&0.67&[  -468,  -461]&  -464&0.67&[  -466,  -460]&  -463&0.67&[  -466,  -462]&  -463\\
NGC 5927 &0.14&[  1099,  1073]&  1084&0.22&[  1315,  1075]&  1193&0.13&[  1079,  1046]&  1066&0.13&[  1083,  1064]&  1074\\
NGC 5946 &0.87&[   147,   126]&   137&0.85&[   147,   130]&   139&0.86&[   147,   137]&   142&0.86&[   152,   132]&   142\\
ES0 224-8 &0.20&[  3261,  3244]&  3253&0.19&[  3253,  3245]&  3249&0.19&[  3249,  3245]&  3247&0.19&[  3248,  3245]&  3247\\
NGC 5986 &0.88&[   104,    85]&    94&0.86&[   105,    89]&    96&0.87&[   108,    93]&   100&0.86&[   117,    92]&   102\\
FSR 1716 &0.34&[  1104,  1061]&  1080&0.31&[  1121,  1075]&  1101&0.31&[  1103,  1076]&  1088&0.31&[  1099,  1071]&  1086\\
BH 184  &0.40&[   568,   511]&   546&0.40&[   548,   522]&   533&0.44&[   591,   485]&   543&0.38&[   537,   515]&   528\\
NGC 6093 &0.81&[    31,    21]&    26&0.81&[    28,    20]&    24&0.81&[    30,    20]&    25&0.78&[    37,    25]&    31\\
NGC 6121 &0.95&[    64,    53]&    59&0.95&[    63,    52]&    59&0.95&[    64,    57]&    60&0.94&[    64,    57]&    61\\
NGC 6144 &0.46&[  -246,  -238]&  -241&0.48&[  -244,  -233]&  -239&0.45&[  -245,  -239]&  -242&0.46&[  -244,  -238]&  -241\\
NGC 6139 &0.51&[   254,   240]&   249&0.48&[   254,   240]&   247&0.49&[   256,   224]&   241&0.48&[   266,   246]&   257\\
Terzan 3 &0.32&[   457,   433]&   447&0.32&[   459,   437]&   448&0.37&[   509,   426]&   470&0.33&[   447,   424]&   436\\
NGC 6171 &0.51&[   196,   184]&   191&0.55&[   195,   183]&   189&0.52&[   200,   177]&   188&0.50&[   206,   184]&   195\\
ESO 452-11 &0.90&[   -19,   -11]&   -15&0.90&[   -18,   -13]&   -15&0.90&[   -18,   -13]&   -15&1.00&[   -17,    -0]&    -2\\
NGC 6218 &0.31&[   591,   569]&   578&0.32&[   591,   556]&   572&0.31&[   611,   574]&   594&0.29&[   602,   580]&   591\\
FSR 1735 &0.95&[   -27,   -14]&   -21&0.93&[   -25,   -18]&   -21&0.94&[   -27,   -16]&   -21&0.92&[   -27,   -14]&   -19\\
NGC 6235 &0.46&[   591,   564]&   577&0.48&[   580,   489]&   536&0.47&[   575,   555]&   566&0.56&[   644,   557]&   600\\
NGC 6254 &0.39&[   611,   590]&   598&0.38&[   611,   577]&   593&0.39&[   636,   602]&   620&0.38&[   620,   602]&   610\\
NGC 6256 &0.92&[    95,    32]&    62&0.89&[    88,    75]&    81&0.90&[   112,    72]&    93&0.88&[    85,    74]&    79\\
NGC 6266 &0.30&[   223,   213]&   218&0.29&[   222,   213]&   218&0.29&[   226,   214]&   220&0.31&[   246,   214]&   230\\
NGC 6273 &0.80&[  -150,  -140]&  -145&0.80&[  -150,  -140]&  -145&0.81&[  -158,  -140]&  -148&0.81&[  -149,  -143]&  -146\\
NGC 6287 &0.94&[   -68,   -55]&   -60&0.93&[   -67,   -50]&   -59&0.93&[   -67,   -57]&   -62&0.93&[   -66,   -56]&   -60\\
NGC 6293 &0.79&[   -98,   -88]&   -93&0.79&[   -96,   -89]&   -93&0.79&[  -108,   -90]&   -98&0.79&[   -96,   -90]&   -93\\
NGC 6304 &0.30&[   488,   472]&   482&0.31&[   490,   472]&   479&0.34&[   531,   470]&   502&0.31&[   478,   437]&   458\\
NGC 6316 &0.75&[   116,   102]&   109&0.74&[   114,   103]&   107&0.74&[   115,   103]&   109&0.75&[   138,    98]&   118\\
NGC 6325 &0.70&[  -120,   -93]&  -105&0.67&[  -109,  -102]&  -105&0.69&[  -108,  -102]&  -105&0.65&[  -109,  -104]&  -107\\
NGC 6333 &0.81&[   327,   310]&   320&0.81&[   328,   318]&   322&0.81&[   331,   317]&   323&0.81&[   331,   320]&   326\\
NGC 6342 &0.53&[   124,   116]&   119&0.51&[   123,   114]&   119&0.52&[   125,   115]&   120&0.54&[   136,   112]&   125\\
\hline
				\end{tabular}\end{center}%			\end{tiny}}
\end{small}
\end{minipage}}

\rotatebox{90}{
	\label{t:A2}
		\begin{minipage}{1.5\linewidth}
		\label{t:A2}
		\begin{small} \begin{center}
{\bf Table A2.}-- Continued. Orbit parameters of globular clusters in barred NFWBB potential
\medskip
\begin{tabular}{|l|ccc|ccc|ccc|ccc|}\hline
Name&\multicolumn{3}{|c|}{$\Omega_{bar}=31$ km s$^{-1}$ kpc$^{-1}$}&\multicolumn{3}{|c|}{$\Omega_{bar}=41$ km s$^{-1}$ kpc$^{-1}$}&
\multicolumn{3}{|c|}{$\Omega_{bar}=55$ km s$^{-1}$ kpc$^{-1}$}&\multicolumn{3}{|c|}{$\Omega_{bar}=70$ km s$^{-1}$ kpc$^{-1}$}\\
 &$e$&$[L_z^{max},L_z^{min}]$&$<L_z>$&$e$&$[L_z^{max},L_z^{min}]$&$<L_z>$&
 $e$&$[L_z^{max},L_z^{min}]$&$<L_z>$&$e$&$[L_z^{max},L_z^{min}]$&$<L_z>$\\
 &  &kpc km s$^{-1}$&kpc km s$^{-1}$&  &kpc km s$^{-1}$&kpc km s$^{-1}$&  &kpc km s$^{-1}$&kpc km s$^{-1}$&  &kpc km s$^{-1}$&kpc km s$^{-1}$\\\hline
NGC 6356 &0.50&[   738,   706]&   721&0.49&[   713,   680]&   697&0.51&[   715,   691]&   704&0.49&[   730,   707]&   719\\
NGC 6355 &0.77&[  -100,   -88]&   -93&0.75&[   -97,   -89]&   -93&0.75&[  -100,   -91]&   -95&0.78&[   -97,   -91]&   -94\\
NGC 6352 &0.23&[   812,   759]&   783&0.18&[   828,   793]&   807&0.24&[   796,   622]&   715&0.16&[   795,   774]&   786\\
Terzan 2 &0.76&[   -49,   -44]&   -46&0.75&[   -49,   -44]&   -46&0.75&[   -48,   -44]&   -46&0.78&[   -49,   -43]&   -46\\
NGC 6366 &0.42&[   714,   690]&   700&0.46&[   740,   684]&   712&0.41&[   707,   676]&   693&0.41&[   703,   685]&   693\\
Terzan 4 &0.39&[   100,    89]&    94&0.35&[    95,    89]&    92&0.35&[    95,    90]&    93&0.35&[    97,    91]&    94\\
BH 229  &0.77&[   -25,   -18]&   -21&0.76&[   -26,   -19]&   -22&0.78&[   -24,   -19]&   -21&0.78&[   -25,   -18]&   -21\\
NGC 6362 &0.31&[   599,   579]&   587&0.33&[   626,   577]&   600&0.32&[   584,   555]&   572&0.32&[   585,   568]&   577\\
Liller1 &0.68&[   -44,   -41]&   -42&0.68&[   -44,   -40]&   -42&0.67&[   -44,   -41]&   -42&0.67&[   -44,   -41]&   -42\\
NGC 6380 &0.72&[  -106,   -96]&  -101&0.72&[  -106,   -99]&  -103&0.82&[  -109,   -93]&  -101&0.74&[  -107,   -99]&  -103\\
Terzan 1 &0.61&[   104,    93]&    99&0.59&[   104,    96]&   100&0.63&[   104,    84]&    95&0.59&[   104,    96]&   100\\
Pismis 26 &0.62&[   305,   267]&   286&0.63&[   287,   269]&   277&0.64&[   279,   245]&   262&0.63&[   275,   262]&   268\\
NGC 6388 &0.46&[  -258,  -250]&  -253&0.45&[  -262,  -253]&  -257&0.47&[  -260,  -253]&  -256&0.46&[  -258,  -254]&  -256\\
NGC 6402 &0.70&[   173,   151]&   162&0.68&[   163,   151]&   157&0.75&[   225,   147]&   187&0.68&[   163,   149]&   156\\
NGC 6401 &0.47&[  -602,  -579]&  -590&0.43&[  -598,  -593]&  -595&0.44&[  -599,  -593]&  -596&0.43&[  -598,  -594]&  -596\\
NGC 6397 &0.40&[   820,   786]&   800&0.40&[   861,   762]&   807&0.37&[   799,   777]&   790&0.38&[   802,   773]&   786\\
Pal 6   &0.91&[    61,    46]&    54&0.94&[    55,    42]&    48&0.93&[    64,    46]&    55&0.93&[    56,    48]&    52\\
Djorg 1  &0.84&[   352,   336]&   345&0.85&[   353,   341]&   347&0.84&[   353,   339]&   346&0.84&[   354,   345]&   349\\
Terzan 5 &0.55&[   113,   101]&   107&0.55&[   111,   103]&   107&0.60&[   111,    86]&    98&0.55&[   111,   102]&   106\\
NGC 6440 &0.68&[   -51,   -47]&   -49&0.69&[   -51,   -47]&   -49&0.68&[   -55,   -43]&   -49&0.69&[   -50,   -46]&   -48\\
NGC 6441 &0.55&[   236,   221]&   230&0.55&[   237,   223]&   228&0.56&[   236,   216]&   226&0.60&[   255,   206]&   231\\
Terzan 6 &0.76&[   -88,   -73]&   -81&0.76&[   -81,   -75]&   -78&0.76&[   -81,   -76]&   -78&0.86&[   -92,   -55]&   -73\\
NGC 6453 &0.75&[   129,    92]&   108&0.73&[   135,   120]&   127&0.73&[   133,    99]&   118&0.74&[   137,   122]&   130\\
NGC 6496 &0.50&[  1154,  1099]&  1125&0.50&[  1125,  1102]&  1110&0.49&[  1134,  1101]&  1119&0.48&[  1117,  1098]&  1107\\
Terzan 9 &0.81&[    40,    26]&    34&0.82&[    34,    27]&    31&0.80&[    31,    27]&    29&0.80&[    31,    27]&    29\\
Djorg 2  &0.54&[   331,   314]&   324&0.54&[   330,   314]&   320&0.56&[   342,   313]&   328&0.60&[   322,   226]&   275\\
NGC 6517 &0.82&[   141,   113]&   130&0.83&[   130,   118]&   124&0.84&[   171,   116]&   146&0.81&[   131,   117]&   124\\
Terzan 10 &0.83&[   204,   187]&   195&0.81&[   201,   184]&   192&0.83&[   196,   173]&   184&0.83&[   215,   187]&   203\\
NGC 6522 &0.56&[    64,    56]&    60&0.57&[    62,    56]&    59&0.57&[    63,    56]&    59&0.57&[    64,    56]&    60\\
NGC 6535 &0.62&[  -320,  -301]&  -312&0.64&[  -324,  -310]&  -316&0.63&[  -323,  -316]&  -319&0.62&[  -321,  -317]&  -319\\
NGC 6528 &0.66&[    53,    43]&    48&0.65&[    57,    46]&    51&0.66&[    54,    45]&    49&0.68&[    57,    40]&    49\\
NGC 6539 &0.32&[   351,   336]&   345&0.24&[   351,   337]&   343&0.26&[   361,   308]&   337&0.23&[   364,   343]&   354\\
NGC 6540 &0.19&[   454,   442]&   448&0.19&[   456,   439]&   446&0.20&[   463,   435]&   448&0.30&[   489,   323]&   402\\
NGC 6544 &0.97&[    44,    25]&    35&0.96&[    37,    23]&    30&0.96&[    36,    26]&    31&0.96&[    35,    28]&    32\\
\hline
				\end{tabular}\end{center}%			\end{tiny}}
\end{small}
\end{minipage}}

\rotatebox{90}{
	\label{t:A2}
		\begin{minipage}{1.5\linewidth}
		\label{t:A2}
		\begin{small} \begin{center}
{\bf Table A2.}-- Continued. Orbit parameters of globular clusters in barred NFWBB potential
\medskip
\begin{tabular}{|l|ccc|ccc|ccc|ccc|}\hline
Name&\multicolumn{3}{|c|}{$\Omega_{bar}=31$ km s$^{-1}$ kpc$^{-1}$}&\multicolumn{3}{|c|}{$\Omega_{bar}=41$ km s$^{-1}$ kpc$^{-1}$}&
\multicolumn{3}{|c|}{$\Omega_{bar}=55$ km s$^{-1}$ kpc$^{-1}$}&\multicolumn{3}{|c|}{$\Omega_{bar}=70$ km s$^{-1}$ kpc$^{-1}$}\\
 &$e$&$[L_z^{max},L_z^{min}]$&$<L_z>$&$e$&$[L_z^{max},L_z^{min}]$&$<L_z>$&
 $e$&$[L_z^{max},L_z^{min}]$&$<L_z>$&$e$&$[L_z^{max},L_z^{min}]$&$<L_z>$\\
 &  &kpc km s$^{-1}$&kpc km s$^{-1}$&  &kpc km s$^{-1}$&kpc km s$^{-1}$&  &kpc km s$^{-1}$&kpc km s$^{-1}$&  &kpc km s$^{-1}$&kpc km s$^{-1}$\\\hline
NGC 6541 &0.56&[   364,   334]&   346&0.56&[   348,   327]&   337&0.59&[   357,   271]&   318&0.56&[   342,   326]&   334\\
NGC 6553 &0.32&[   592,   571]&   584&0.31&[   602,   580]&   590&0.41&[   740,   554]&   650&0.30&[   603,   573]&   589\\
NGC 6558 &0.77&[    97,    85]&    92&0.77&[    98,    84]&    90&0.75&[    98,    86]&    92&0.82&[   127,    86]&   108\\
Pal 7   &0.32&[  1046,  1004]&  1021&0.33&[  1043,   932]&  1000&0.31&[  1055,  1028]&  1045&0.31&[  1053,  1019]&  1035\\
Terzan 12 &0.32&[   680,   606]&   641&0.32&[   629,   601]&   613&0.35&[   702,   610]&   657&0.32&[   641,   620]&   632\\
NGC 6569 &0.20&[   444,   432]&   438&0.19&[   447,   432]&   438&0.21&[   459,   420]&   439&0.20&[   462,   423]&   444\\
ESO 456-78 &0.41&[   380,   364]&   374&0.40&[   386,   369]&   376&0.42&[   409,   359]&   385&0.40&[   384,   354]&   370\\
NGC 6624 &0.39&[    38,    35]&    36&0.39&[    37,    35]&    36&0.39&[    37,    34]&    36&0.40&[    38,    34]&    36\\
NGC 6626 &0.61&[   175,   163]&   170&0.62&[   177,   161]&   167&0.62&[   175,   161]&   167&0.69&[   226,   138]&   179\\
NGC 6638 &0.85&[    24,    15]&    19&0.86&[    28,    17]&    23&0.89&[    25,    17]&    21&0.87&[    25,    18]&    21\\
NGC 6637 &0.44&[    44,    40]&    42&0.46&[    44,    39]&    41&0.47&[    44,    40]&    42&0.51&[    49,    39]&    44\\
NGC 6642 &0.85&[    42,    26]&    34&0.87&[    39,    31]&    35&0.87&[    40,    32]&    36&0.87&[    38,    30]&    34\\
NGC 6652 &0.85&[    47,    36]&    43&0.86&[    49,    39]&    42&0.85&[    47,    34]&    40&0.84&[    52,    41]&    47\\
NGC 6656 &0.56&[  1053,   941]&   994&0.56&[  1055,  1033]&  1045&0.55&[  1056,   983]&  1018&0.53&[  1048,  1032]&  1040\\
Pal 8   &0.43&[   616,   589]&   600&0.44&[   621,   587]&   603&0.44&[   594,   513]&   550&0.44&[   593,   571]&   583\\
NGC 6681 &0.94&[    41,    27]&    34&0.94&[    43,    25]&    34&0.94&[    52,    23]&    37&0.88&[    38,    31]&    34\\
NGC 6712 &0.89&[   108,    91]&    99&0.89&[   100,    86]&    94&0.86&[   102,    89]&    96&0.85&[    98,    90]&    94\\
NGC 6717 &0.31&[   257,   246]&   251&0.27&[   251,   237]&   245&0.26&[   254,   239]&   246&0.30&[   254,   211]&   234\\
NGC 6723 &0.78&[   -12,     0]&    -4&0.94&[    -7,     0]&    -1&1.00&[     9,    -0]&     0&0.90&[   -25,    -0]&    -2\\
NGC 6749 &0.46&[   701,   555]&   634&0.46&[   567,   541]&   553&0.53&[   618,   474]&   553&0.46&[   570,   549]&   561\\
NGC 6752 &0.20&[   946,   920]&   932&0.26&[  1113,   905]&  1006&0.19&[   938,   912]&   925&0.19&[   944,   921]&   933\\
NGC 6760 &0.40&[   743,   712]&   724&0.40&[   733,   697]&   713&0.41&[   763,   722]&   744&0.40&[   746,   718]&   727\\
Pal 10  &0.25&[  1248,  1209]&  1225&0.27&[  1315,  1221]&  1272&0.25&[  1263,  1231]&  1244&0.26&[  1263,  1182]&  1223\\
NGC 6809 &0.71&[   268,   251]&   260&0.70&[   275,   242]&   260&0.72&[   274,   258]&   266&0.71&[   272,   259]&   266\\
Pal 11  &0.39&[  1062,  1006]&  1032&0.37&[  1015,   982]&  1001&0.37&[  1018,   993]&  1005&0.37&[  1025,  1008]&  1018\\
NGC 6838 &0.17&[  1443,  1390]&  1411&0.17&[  1457,  1403]&  1435&0.16&[  1440,  1417]&  1428&0.16&[  1434,  1418]&  1427\\
NGC 7078 &0.48&[  1155,  1109]&  1133&0.47&[  1138,  1118]&  1127&0.48&[  1122,  1087]&  1107&0.47&[  1122,  1100]&  1111\\
\hline
				\end{tabular}\end{center}%			\end{tiny}}
\end{small}
\end{minipage}}

 \end{document}